\documentclass[epj]{svjour}
\usepackage{latexsym}
\usepackage{amssymb,amsmath}
\usepackage{bm,epsf}
\usepackage{color}

\newcommand{\doo}[2]{\ensuremath{{\partial #1\over \partial #2}}}

\def\mv{{\mathbf v}}

\def\mx{{\mathbf x}}

\def\my{{\mathbf y}}

\newcommand{\dee}[2]{\ensuremath{{{\rm d} #1\over {\rm d} #2}}}

\usepackage{graphics}
\begin{document}
\title{Two-loop calculation of anomalous kinetics of the reaction $A+A\rightarrow\varnothing$ in randomly stirred fluid}
\titlerunning{Two-loop calculation of the reaction $A+A\rightarrow\varnothing$.}
\author{Michal Hnati\v{c}\inst{1,2} \and Juha Honkonen\inst{3} and Tom\'a\v{s} Lu\v{c}ivjansk\'y\inst{1,2} 
}                     
\institute{Institute of Experimental Physics, SAS, Ko\v{s}ice, Slovakia
          \and Faculty of Sciences, P.J. \v{S}afarik University, Ko\v{s}ice, Slovakia
          \and Department of Military Technology, National Defence University, Helsinki, Finland}
\date{Received: date / Revised version: date}
\authorrunning{M. Hnati\v{c}, J. Honkonen,T. Lu\v{c}ivjansk\'y}

\abstract{
The single-species annihilation reaction $\textit{A}
+\textit{A} \rightarrow \varnothing$ is studied in the presence of
a random velocity field generated by the stochastic Navier-Stokes
equation. The renormalization group is used
to analyze the combined influence of the density and velocity
fluctuations on the long-time behavior of the
system. The direct effect of velocity fluctuations on the
reaction constant appears only from the two-loop order, therefore
all stable fixed points of the renormalization group and their regions of stability are
calculated in the two-loop approximation in the two-parameter $(\epsilon,\Delta)$
expansion. A renormalized integro-differential equation for
the number density is put forward which takes into account the
effect of density and velocity fluctuations at next-to-leading order.
Solution of this equation in perturbation theory is calculated in a homogeneous system.}

\maketitle

\section{Introduction}
\label{intro}
The irreversible annihilation reaction $\textit{A} +\textit{A} \rightarrow
\varnothing$ is a fundamental model of non-equilibrium physics.
The reacting
$A$  particles  are assumed to
 perform chaotic motion due to diffusion or some
external advection field such as atmospheric eddy and may react after the mutual collision with
constant microscopic probability $K_0$ per unit time.
Usually the resulting molecule is considered to be
chemically inert
with no backward
influence on the movement of the reacting $A$ particles.

Many reactions of this type are observed in diverse
chemical, biological or physical systems \cite{Derrida95,Kroon93}.
The usual approach
to such
kind of problems is based on the use of the kinetic rate equation.
It leads to a self-consistent description analogous to the mean-field approximation in the
theory of critical phenomena. Its basic assumption  is that the particle
density is spatially homogeneous
(fluctuations in concentration field are neglected).
 This homogeneity can be thought as a consequence of
either an infinite mobility of the reactants or of a very small probability
that a chemical reaction actually occurs when reacting entities meet each other.
On the other hand, if the particle mobility becomes sufficiently small, or
equivalently, if the microscopic reaction probability becomes
large enough there is a possible transition to a new regime where it is more
probable that the given particle reacts with local neighbors than with distant
particles. This behavior is known as the diffusion-controlled regime \cite{Kampen,Kang84}.
For the annihilation reaction limited assumption of the density homogeneity leads to the following
equation for the mean number density
\begin{equation}
   \partial_t n(t) = -K_0 n^2(t).
  \label{eq:mpn}
\end{equation}
This equation predicts a long-time asymptotic decay as $n(t)\sim t^{-1}$ and the decay exponent does not depend on
the space dimension. This is a common situation observed in the mean field theory.
 However, it turns out \cite{Lee94,Cardy96} that in lower space dimensions $d\leq 2$ the assumption of
spatially uniform density, or equivalently of negligible density fluctuations of reacting particles, is not appropriate.
Reentrant property of the diffusing particles \cite{Itzykson} in low space dimensions leads to
effective slowing-down of the reaction process and it can be
rigorously shown, that
the upper critical dimension for this process
is
$d_c=2$ \cite{Lee94}, above which the mean field approximation is valid.

A typical reaction occurs in liquid or gaseous environment.
Thermal fluctuations of this underlying environment cause
additional advection of the reacting particles. Therefore, it is
interesting to study the influence of the advection field
on the annihilation process.

Most of the
renormalization-group analyses of the effect of random drift
on the annihilation reaction  $\textit{A} +\textit{A} \rightarrow \varnothing$ in the framework
of the Doi approach
have been carried out for the case of a quenched random drift field. Potential random drift with long-range \cite{Park,Chung}
and short-range correlations \cite{Richardson99} have been studied as well as ''turbulent'' flow (i.e. quenched solenoidal random field) with potential
disorder \cite{Deem2,Tran}. For a more realistic description of a turbulent flow time-dependent velocity field would be more appropriate.
In Ref. \cite{Tran} dynamic disorder with a given Gaussian distribution has been considered, whereas the most ambitious approach
on the basis of a velocity field generated by the stochastic Navier-Stokes equation has been introduced here by two of the present authors  \cite{Hnatic}.
From the point of the Navier-Stokes equation the situation near the critical dimension $d_c=2$ of the pure reaction model is even more intriguing
due to the properties of the Navier-Stokes equation.
It is well-known fact \cite{Frisch} that in the case of space dimension $d=2$, there is inviscid
conservation law of  enstrophy absent in the three dimensional case.
Calculations in Ref. \cite{Hnatic} were performed in the one-loop
approximation. As may readily seen from examination of the Feynman graphs,
in the one-loop approximation there is no influence of the velocity fluctuations on the
renormalization of the interaction vertices. However, the influence of higher
order terms of the perturbation series can have significant effect on the critical properties.
In this paper we study the advection of reactive scalar using random velocity field generated
by the stochastic Navier-Stokes equation, which is used for production a velocity field
corresponding to thermal fluctuations \cite{Forster1,Forster2} and a turbulent velocity
field with the Kolmogorov scaling behavior \cite{Adzhemyan}.
 It should be stressed, moreover, that in the presented model we assume that
 there is no influence of the reactant on the velocity field itself.
Therefore, the model may be characterized as a model for the advection of a passive chemically active admixture.

A powerful tool for
analyzing the asymptotic behavior of stochastic systems is provided by the renormalization-group (RG) method.
It allows to determine the long-time and large-scale -- or infra-red (IR) -- asymptotic regimes of
the system and also is very efficient tool for calculation of various universal
physical quantities, e.g. critical exponents.
The aim of this paper is to examine the IR behavior of the
annihilation process under the influence of advecting velocity fluctuations and to determine its
stability in the second order of the perturbation theory.

Using the mapping procedure based on the Doi formalism
\cite{Doi1,Doi2} an effective field-theoretic model for the annihilation process is constructed.
The RG method is applied to the model
in the field-theoretic formulation, which is the most efficient in calculations beyond
the one-loop order,
and the renormalization
constants and fixed points of the renormalization group are determined in the two-loop approximation within the two-parameter expansion.
The non-linear integro-differential equation, which includes first non-trivial
corrections to (\ref{eq:mpn}), is obtained for the mean number density and it is
shown how the information about IR asymptotics can be extracted from it
in the case of a homogeneous system. This equation
allows to investigate heterogeneous systems as well as to take into account the effect
of density and advecting velocity fluctuations. However, solution of the equation
in the heterogenous case requires heavy numerical calculations and is beyond the
scope of the present paper. We intend to return to this problem in future work.

The paper is organized as follows. In Sec. II the field-theoretic model for the
annihilation process is constructed on the basis of the second-quantization approach.
The basic ingredients for the modeling of a velocity field by the stochastic Navier-Stokes equation are presented.
It is also shown how both the Kolmogorov scaling and thermal fluctuations can be included into the model.
The ultraviolet (UV) renormalization of the model and the elaborated algorithm for
the calculation of the renormalization constants is described in Sec. III. Fixed points
of the RG are classified together with their stability regions and possible scaling regimes are presented in Sec IV.
In Sec. V the integro-differential equation for the mean number density is derived and
analysis of its solution is given. Conclusions are presented in Sec. VI.
\section{Field-theoretic model of the annihilation reaction}
Let us study anomalous kinetics of the generic type of the irreversible single-species
annihilation reaction
\begin{equation}
  A+A \xrightarrow{K_0} \varnothing,
  \label{eq:rea}
\end{equation}
with the unrenormalized (mean field) rate constant $K_0$.
The first step of the Doi approach \cite{Doi1,Doi2} (see also \cite{Peliti}) consists of the introduction of
the creation and annihilation
operators $\psi^\dagger$ and $\psi$ and the vacuum state $|0\rangle$
satisfying the usual bosonic commutation relations
\begin{eqnarray}
 & & [\psi(\mathbf{x}),\psi^\dagger({\mathbf x'})]=\delta(\mathbf{x}-{\mathbf x'}),\nonumber\\*
 & &[\psi(\mathbf{x}),\psi({\mathbf x'})]=[\psi^\dagger(\mathbf{x}),\psi^\dagger(\mathbf{x'})] = 0, \nonumber\\*
 & & \psi(\mathbf{x})|0\rangle   = 0, \langle 0| \psi^\dagger(\mathbf{x}) = 0, \langle 0|0 \rangle = 1.
  \label{eq:comm_rel}
\end{eqnarray}
Let $P(\{n_i\},t)$ be the joint probability density function (PDF) for observing $n_i$ particles at positions
$\mathbf{x}_i$. The information about the macroscopic state of the classical many-particle system may be
 transferred into the state vector $ | \Phi(t)\rangle$ defined as the sum over all occupation numbers
\begin{equation}
 | \Phi(t)\rangle = \sum_{ \{n_{i}\} } P(\{ n_i \},t) | \{ n_i \} \rangle,
 \label{eq:sta_vec}
\end{equation}
where the basis vectors are defined as
\begin{equation}
 | \{ n_i \} \rangle = \prod_{i} [\psi^\dagger(\mathbf{x}_i) ]^{n_i} |0\rangle.
 \label{eq:blabol}
\end{equation}
The whole set of coupled partial differential equations for the PDFs may be
 rewritten in the compact form  of a master equation \cite{Lee94,Doi1,Doi2}
\begin{equation}
 \frac{\partial}{\partial t}| \Phi(t)\rangle = -\hat{H} | \Phi(t) \rangle,
 \label{eq:master}
\end{equation}
where $\hat{H} = \hat{H}_A+\hat{H}_D+\hat{H}_R$ and for the annihilation process $A+A \rightarrow \varnothing$
under consideration
\begin{eqnarray}
   & &\hat{H}_A = \int d{\mathbf x}\mbox{ } \psi^\dagger \nabla [{\mathbf v}({\mathbf x},t) \psi({\mathbf x})],\nonumber\\*
   & &\hat{H}_D = - D_{0} \int d{\mathbf x} \mbox{ } \psi^\dagger \nabla^{2} \psi({\mathbf x}),\nonumber\\*
   & &\hat{H}_R = \lambda_{0}D_0 \int d{\mathbf x}\mbox{ } (\psi^\dagger)^{2} \psi^{2},
   \label{eq:hamil_react}
\end{eqnarray}
corresponding to the advection, diffusion and reaction parts of the operator $\hat{H}$ \cite{Hnatic}.
Due to dimensional reasons we have extracted
the diffusion constant $D_0$ from the rate constant $K_0=\lambda_0 D_0$.

The mean of an observable quantity $\mathcal{O}(t)$ may be expressed \cite{Hnatic} as the vacuum expectation value
\begin{eqnarray}
   \langle \mathcal{O}(t) \rangle  & = &  \langle 0| T(\mathcal{O}\{[\psi^\dagger(t)+1]\psi(t) \}  \nonumber\\*
    & \times & {\rm e}^{-\int^{\infty}_{0} \hat{H}'_{I} dt + n_{0} \int d{\mathbf x}\mbox{ }
     \psi^\dagger({\mathbf x},0)}  )|0\rangle.
  \label{eq:aver2}
\end{eqnarray}
Here, the interaction operator is defined as $\hat{H}'_I=\hat{H}'-\hat{H}'_0$
and the substitution $\psi^\dagger\rightarrow\psi^\dagger+1: \hat{H}'\equiv \hat{H}(\psi^\dagger+1,\psi)$
is understood. The field operators 
 (\ref{eq:comm_rel}) are replaced by the time-dependent operators
 of the interaction representation
\begin{equation*}
  \psi^\dagger(t,{\mathbf x})={\mathrm e}^{\hat{H}_0't}\psi^\dagger({\mathbf x}){\mathrm e}^{-\hat{H}_0't},\quad
   \psi(t,{\mathbf x})={\mathrm e}^{\hat{H}_0't} \psi({\mathbf x}){\mathrm e}^{-\hat{H}_0't}.
\end{equation*}
In this formulation - assuming the Poisson distribution
as the initial condition - the average number density can be computed via the expression
\begin{equation}
   n(t,{\mathbf x}) = \langle 0| \psi({\mathbf x}){\mathrm e}^{-\hat{H}' t}
   {\mathrm e}^{n_0\int d{\mathbf x}\psi^\dagger}|0\rangle,
   \label{eq:aver_number}
\end{equation}
where $n_0$ is the initial number density.
The expectation value of the time-ordered product in (\ref{eq:aver2}) can
be cast \cite{Vasiliev} into the form of a
functional integral over scalar fields
$\psi^\dagger(\mathbf{x},t)$ and $\psi(\mathbf{x},t)$:
\begin{equation}
 \langle \mathcal{O}(t) \rangle =
 \int \mathcal{D}\psi^\dagger \mathcal{D}\psi \mbox{ } \mathcal{O}\{ [\psi^\dagger(t)+1]\psi(t)\} \rm{e}^{S_1},
  \label{eq:expect_val}
\end{equation}
where the action $S_{1}$ for the annihilation reaction $\textit{A} +\textit{A} \rightarrow \varnothing$  is
\begin{eqnarray}
  S_{1}  & = &  - \int^{\infty}_{0} dt \int d{\mathbf x} \{ \psi^\dagger \partial_{t} \psi +
  \psi^\dagger\nabla({\mathbf v}\psi) -D_{0}\psi^\dagger\nabla^{2}\psi
   \nonumber\\*
  & + & \lambda_{0} D_{0}[2\psi^\dagger+(\psi^\dagger)^{2}]\psi^{2}  \} +
  n_{0}\int d{\mathbf x}\mbox{ } \psi^\dagger({\mathbf x},0) .
  \label{eq:S_1}
\end{eqnarray}
In order to analyze the effect of velocity fluctuations on the reaction process
we average the expectation value (\ref{eq:expect_val}) over
the random velocity field $\mathbf{v}$. The most realistic description of the velocity field
${\bf v}$ is based on the use of the stochastic Navier-Stokes equation
\cite{Adzhemyan}.
Due to the incompressibility conditions $\nabla\cdot\mathbf{v}=0$ and $\nabla\cdot\mathbf{f}^v=0$ imposed
on the velocity field $\mathbf{v}$ and the random-force field $\mathbf{f}^v$ it is possible
to eliminate pressure from the Navier-Stokes equation and hence it is sufficient to consider only
its
transverse components
\begin{equation}
  \partial_{t}{\mathbf v} + P({\mathbf v}\cdot\nabla){\mathbf v}-\nu_{0}\nabla^{2}{\mathbf v}
  ={\mathbf f}^{v}.
  \label{eq:NS}
\end{equation}
 Here, $\nu_0$ is the molecular kinematic viscosity,
$P_{ij}({\bf k}) = \delta_{ij}-k_ik_j/k^2$ is the transverse
projection operator and $k=|{\bf k}|$ is the norm of the wave vector ${\mathbf k}$.
Here and below we use the subscript $"0"$ for all "bare" parameters
to distinguish them from their renormalized counterparts, which will appear during
the renormalization procedure.

The large-scale random force per unit mass ${\mathbf f}$ is assumed to be a Gaussian random variable with zero mean and the following correlation function
\begin{eqnarray}
 & & \langle f_{m}({\mathbf x}_1,t_1) f_{n}({\mathbf x}_2,t_2) \rangle =  \delta (t_1-t_2) \nonumber\\*
 &  & \times\int \frac{d{\mathbf k}}{(2\pi)^d} P_{mn}({\mathbf k}) d_f(k)
  \rm{e}^{i{\mathbf k}\cdot({\mathbf x}_{1}-{\mathbf x}_{2}  )},
  \label{eq:cor_ran_for}
\end{eqnarray}
where the kernel function is chosen in the form
\begin{equation}
 d_f(k) =  g_{10} \nu_{0}^{3}  k^{4-d-2\epsilon}  + g_{20}\nu_{0}^{3}k^{2}.
  \label{eq:kernel}
\end{equation}
The nonlocal term is often used to generate the turbulent velocity field with Kolmogorov's scaling
\cite{Adzhemyan,Dominicis,Vasiliev04}. That case
is achieved by setting $\epsilon=2$. The local term $g_{20}\nu_0^3 k^2$ has been added not only because of renormalization reasons but has also
an important physical meaning. Such a term in the force correlation function describes generation of thermal fluctuations
of the velocity field near equilibrium \cite{Forster1,Forster2} and thus can mimic the usual environment in which
chemical reactions take place.

Averaging
in the expectation value
 ($\ref{eq:expect_val}$)
over the realizations of
the random velocity field ${\mathbf v}$  is done with the use of the
"weight" functional $\mathcal{W}_{2} = \rm{e}^{S_{2}}$.
Here, $S_{2}$ is the effective action for the advecting velocity field
\cite{Vasiliev04}
\begin{eqnarray}
   S_{2} & = & \frac{1}{2} \int dt d{\mathbf x} d {\mathbf {x}'}\mbox{ } \tilde{\mathbf v}
  ({\mathbf x},t) \cdot \tilde{ {\mathbf v} } ( {\mathbf x}',t)
   d_f (| {\mathbf x} -{\mathbf x}' |)
  \nonumber\\*
  & + &  \int dt d{\mathbf x} \mbox{ }
   \tilde{{\mathbf v}} \cdot [-\partial_{t} {\mathbf v}-( {\mathbf v}\cdot\nabla)
  {\mathbf v}+\nu_{0}\nabla^{2}{\mathbf v}],
  \label{eq:S_2}
\end{eqnarray}
where $\tilde{{\mathbf v}}$ is the auxiliary transverse vector field, that results
from the Gaussian averaging with respect to the realizations of random force ${\mathbf f}^{v}$.
The appearance of such field is common for the models of stochastic dynamics and the field
itself can
be understood as a response field \cite{Jan76}.

With the use of the complete weight functional
\begin{equation}
\mathcal{W} = \rm{e}^{S_{1} + S_{2}}
\label{eq:total_func}
\end{equation}
it is possible to evaluate the expectation value of any desirable physical observable.

\begin{figure}
  \resizebox{0.7\columnwidth}{!}{%
  \includegraphics{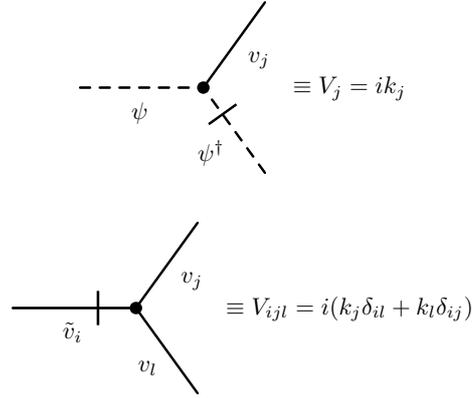}
  }
  \caption{Interaction vertices describing velocity fluctuation and advection
  and the corresponding vertex factors}
  \label{fig:ver_vel}       
\end{figure}

\begin{figure}
  \resizebox{0.7\columnwidth}{!}{%
  \includegraphics{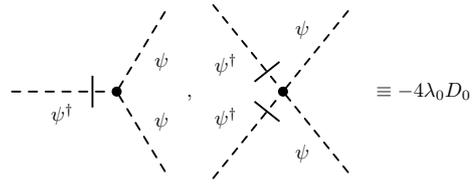}
  }
  \caption{Interaction vertices responsible for density fluctuations and their corresponding vertex factor}
  \label{fig:ver_react}       
\end{figure}

Actions  (\ref{eq:S_1}) and (\ref{eq:S_2}) for the studied model are written in the form
convenient for the use of the standard Feynman diagrammatic technique.
\begin{figure}
  \resizebox{0.75\columnwidth}{!}{%
  \includegraphics{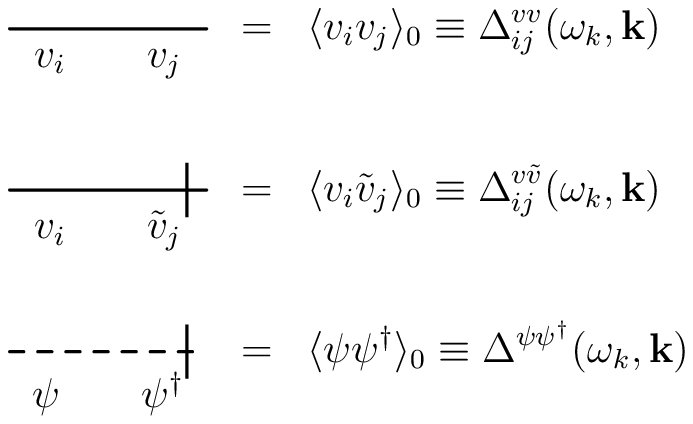}
  }
  \caption{The propagators of the model}
  \label{fig:propag}       
\end{figure}
The inverse matrix of the quadratic part of the actions determines the form of the bare propagators.
In the Feynman graphs these propagators correspond to lines connecting interaction vertices. It is
 easily seen that the studied model
 contains three different types of propagators.
The propagators are presented in the wave-number-frequency representation, which is
the most convenient way for doing explicit calculations. The graphical representation of the propagators
is presented in Fig. \ref{fig:propag}, where
\begin{eqnarray}
 & &   \Delta_{ij}^{v\tilde{v}}(\omega_k,{\mathbf k})= \frac{P_{ij}({\mathbf k})}{-i\omega_k+\nu_0 k^2}\,,
  \Delta_{ij}^{vv}(\omega_k,{\mathbf k})= \frac{d_f(k)P_{ij}^k({\mathbf k})}{\omega_k^2+\nu_0^2k^4}, \nonumber\\*
 & & \hskip2cm
 \Delta^{\psi\psi^\dagger}(\omega_k,{\mathbf k}) = \frac{1}{-i\omega_k+D_0k^2},
   \label{eq:prop}
\end{eqnarray}
with the kernel function $d_f(k)$ given by the expression (\ref{eq:kernel}).
The vertex factor
\begin{equation}
  V_m(x_1,x_2,\ldots,x_m;\Phi) = \frac{\delta^m V(\Phi)}{\delta\Phi(x_1)\delta\Phi(x_2)\ldots\delta\Phi(x_m)}
  \label{eq:ver_factor}
\end{equation}
 is associated to each interaction vertex of Feynman graph. Here, $\Phi$ could be any member from the set of all fields $\{\psi^\dagger,\psi,\tilde{v},v \}$.
The interaction vertices from action (\ref{eq:S_2}) describe
 the advection of reactant particles by the velocity field and the interactions between
 the velocity components.
The interaction vertex in (\ref{eq:S_2}) may be rewritten in a technically more convenient
form
\begin{eqnarray}
  &  & -\int dtd{\mathbf x}\mbox{ } \tilde{\mv}(\mv\cdot\nabla){\mv} =
  - \int dtd{\mathbf x}\mbox{ }\tilde{v}_i v_k\partial_k v_i \nonumber\\*
  & & = \int dtd{\mathbf x}\mbox{ }(\partial_k \tilde{v}_i) v_k v_i,
  \label{eq:int_ver}
\end{eqnarray}
where the incompressibility condition $\partial_i v_i=0$ and partial integration method have been used.
We
have assumed
that the velocity fields fall off rapidly for $|\mathbf x|\rightarrow\infty$
and therefore the surface terms can be neglected.
Rewriting
(\ref{eq:int_ver}) into
 the symmetric form $v_i V_{ijl}v_j v_l/2$, it is
easy to find the explicit form for the corresponding
vertex factor in the momentum space
\begin{equation}
  V_{ijl}   = i(k_j\delta_{il}+k_l\delta_{ij}).
  \label{eq:int_ver_ns}
\end{equation}
Here, the momentum ${\mathbf k}$ is flowing into the vertex through the field $\tilde{v}$. The advecting
term from the action (\ref{eq:S_1}) can be similarly
modified as follows
\begin{eqnarray}
 & &  -\int dtd{\mathbf x}\mbox{ }\psi^\dagger \nabla (\mv \psi)  =
   -\int dtd{\mathbf x}\mbox{ } \psi^\dagger \partial_i(v_i\psi)\nonumber\\*
& &  = -\int dtd{\mathbf x}\mbox{ }\psi^\dagger v_i\partial_i\psi=
\int dtd{\mathbf x}\mbox{ }(\partial_i\psi^\dagger) v_i\psi.
  \label{eq:int_ver_react}
\end{eqnarray}
Rewriting this expression in the form $\psi^\dagger V_j v_j\psi $ we obtain immediately the vertex factor in the
momentum space
\begin{equation}
  V_j=ik_j.
  \label{eq:int_ver_adv}
\end{equation}
The momentum ${\mathbf k}$ represents the momentum flowing into the diagram through the slashed field $\psi^\dagger$.
These two vertices (\ref{eq:int_ver_ns}), (\ref{eq:int_ver_adv})
are depicted in Fig. \ref{fig:ver_vel}.
\mbox{ }
The two reaction vertices derived from the functional (\ref{eq:S_1}) are depicted in
Fig. \ref{fig:ver_react}
and physically describe the density fluctuations of the reactant particles.
The vertex factors for both of them follows from the straightforward application of the definition (\ref{eq:ver_factor}).

\section{UV renormalization of the model}
The functional formulation provides a theoretical framework
suitable for applying methods of quantum field theory.
Using RG methods it is possible to determine the IR
asymptotic (large spatial and time scales) behavior of the correlation functions. First of all,
a proper renormalization procedure is needed for the elimination of ultraviolet (UV) divergences.
There are various renormalization prescriptions applicable for such a task, each with its own
advantages. To most popular belong the Pauli-Villars, lattice
and dimensional regularization \cite{Zinn-Justin89}. In what follows we will employ the
modified minimal subtraction ($\overline{\text{MS}}$) scheme. Strictly speaking, in the analytic
renormalization there is no consistent MS scheme. What we mean here, is the ray scheme \cite{Adzhemyan3}, in
which the two regularizing parameters $\epsilon$, $\Delta$ ($\epsilon$ has been introduced in (\ref{eq:kernel}) and $2\Delta=d-2$) are taken proportional
to each other: $\Delta=\xi \epsilon$, where the coefficient $\xi$ is arbitrary but fixed. In this case,
only one independent small parameter, say, $\epsilon$ remains and the notion of minimal subtraction becomes meaningful.
UV divergences manifest themselves
in the form of poles in the small expansion parameter and the minimal subtraction scheme
is characterized by discarding all finite parts of the Feynman graphs in the calculation of the renormalization
constants.
In the modified scheme, as usual, certain geometric factors are not expanded in $\epsilon$, however.
This is the content of the $\overline{\text{MS}}$ scheme used in our analysis.

In order to apply the dimensional regularization for the evaluation 
of renormalization constants, an analysis of possible superficial
divergences has to be performed. 
For the power counting in the actions (\ref{eq:S_1}) and
(\ref{eq:S_2}) we use the scheme \cite{Adzhemyan}, in which to each
quantity $Q$ two canonical dimensions are assigned, one with respect
to the wave number $d_Q^k$ and the other to the frequency
$d_Q^\omega$. The normalization for these dimensions is
\begin{equation}
 d^\omega_\omega=-d_t^\omega=1,\quad d_k^k=-d_x^k=1,\quad d_k^\omega=d^k_\omega=0.
\label{eq:def_power}
\end{equation}
 The canonical (engineering) dimensions
for fields and parameters of the model are derived from the condition for action
to be a scale-invariant quantity, i.e. to have a zero canonical dimension.
The quadratic part of the action (\ref{eq:S_1}) determines only the
canonical dimension of the quadratic product $\psi^\dagger\psi$. In
order to keep both terms in the nonlinear part of the action
\begin{equation}
\lambda_0 D_0 \int
dtd\mathbf{x}[2\psi^\dagger+(\psi^\dagger)^2]\psi^2,
\label{eq:quad_part}
\end{equation}
the field $\psi^\dagger$ must be dimensionless. If
the field $\psi^\dagger$ has a positive canonical dimension, which
is the case for $d>2$, then
the quartic term should be discarded as irrelevant by the power
counting. The action with the cubic term only, however, does not
generate any loop integrals
corresponding to the density fluctuations and thus is uninteresting for the analysis of fluctuation effects in the
space dimension $d=2$.

Using the normalization choice (\ref{eq:def_power}),
we are able to obtain the canonical dimensions for all the fields and parameters in the
$d$-dimensional space. The results
are summarized in Table \ref{tab:canon_dims}.
\begin{table}
  \caption{Canonical dimensions for the parameters and the fields of the model}
  \label{tab:canon_dims}
  \begin{tabular}{|c|c|c|c|c|c|c|c|c|c|}
    \hline\noalign{\smallskip}{\baselinestretch}{1}
    $Q$ & $\psi$ & $\psi^\dagger$ & $v$ & $\tilde{v}$ & $\nu_0,D_0$ & $\lambda_0$ & $g_{10}$ & $g_{20}$ \\
    \noalign{\smallskip}\hline\noalign{\smallskip}
    $d_Q^\omega $& $0 $ & $0$ & $1$ & $-1$ & $1$ & $0$ & $0$ & $0$\\ \hline
    $d_Q^k$ & $d$ & $0$ & $-1$ & $d+1$ & $-2$ & $-2\Delta$ & $2\epsilon$ & $-2\Delta$ \\ \hline
    $d_Q $& $d$ & $0$ & $1$ & $d-1$ &  $0$  & $-2\Delta$ & $2\epsilon$ & $-2\Delta$\\
    \noalign{\smallskip}\hline
  \end{tabular}
\end{table}
Here, $d_Q=d^k_Q+2d_Q^\omega$ is the total canonical dimension and it is determined from the condition that the parabolic
differential operator of the diffusion and Navier-Stokes equation scales uniformly under the simultaneous momentum and frequency dilatation
$k\rightarrow \mu k, \omega\rightarrow \mu^2\omega$.

The model is logarithmic when
canonical dimensions of 
all the coupling constants $\{g_{10},g_{20},\lambda_0,u_0\}$ vanish simultaneously. From
Table \ref{tab:canon_dims} it follows that this situation occurs for the choice $\epsilon=\Delta=0$.
The parameter $\epsilon$ characterizes the deviation from the Kolmogorov
scaling \cite{Adzhemyan99} observed in the real turbulence and together with $\Delta$ may be considered as
the analog of the expansion parameter $\epsilon=4-d$ used in the theory of critical phenomena. The UV divergences
have the form of poles in various
linear combinations of $\epsilon$ and $\Delta$.
The total canonical dimension of an arbitrary one-particle irreducible Green (1PI)
function $\Gamma=\langle\Phi\ldots \Phi  \rangle_{\rm 1-ir} $ is given by the relation
$  d_\Gamma= d + 2 - N_\Phi d_\Phi,$
where $N_\Phi=\{N_{\psi^\dagger},N_\psi,N_v,N_{\tilde{v}} \}$ are the numbers of corresponding external fields
entering into the function $\Gamma$.

The statistical averaging $\langle\ldots\rangle$ means averaging over all possible realizations
of fields $\tilde{\mv},\mv,\psi^\dagger,\psi$ satisfying appropriate boundary conditions
with the use of the complete weight functional (\ref{eq:total_func}).
Superficial UV divergences may be present
only in those $\Gamma$ functions for which $d_\Gamma$ is a non-negative integer.
Using the dimensions of the fields from Table \ref{tab:canon_dims}
we see that the superficial degree of divergence for a 1PI function $\Gamma$ is given by
the expression
\begin{equation}
d_\Gamma=4-N_v-N_{\tilde{v}}-2N_\psi.
\end{equation}
However, the real degree of divergence $\delta_\Gamma$ is smaller, because of the
structure of the interaction vertex (\ref{eq:int_ver}), which allows for factoring out
the operator $\partial$ to each external line $\tilde{v}$.
Thus
the real divergence exponent $\delta_\Gamma$
may be expressed as
\begin{equation}
  \delta_\Gamma\equiv d_\Gamma-N_{\tilde{v}} = 4-N_v-2N_{\tilde{v}}-2N_\psi
  \label{eq:real_div_exp}
\end{equation}
Although the canonical dimension for
the field $\psi^\dagger$ is zero, there is no proliferation of superficial divergent
graphs with arbitrary number of external $\psi^\dagger$ legs. This is due to the
condition
 $n_{\psi^\dagger}\leq n_\psi$, which
may be established by a straightforward analysis of the
Feynman
graphs \cite{Lee94}.
As has already been shown \cite{Adzhemyan2} the divergences in (1PI) Green functions containing at least one velocity field ${\bf v}$
may be removed by a single counterterm of the form $\psi^\dagger \partial^2 \psi$.

Brief analysis shows that the UV divergences are expected only for the 1PI Green functions listed in
Table \ref{tab:canon_green}.
\begin{table}
  \caption{Canonical dimensions for the (1PI) divergent Green functions of the model}
  \label{tab:canon_green}
  \begin{tabular}{|c|c|c|c|c|c|c|c|c|c|}
    \hline\noalign{\smallskip}
     $\Gamma_{1-ir}$ & $\langle\psi_\dagger \psi\rangle $& $\langle\psi_\dagger\psi v\rangle$
                  & $\langle \tilde{v} v\rangle $ & $ \langle \tilde{v} v v\rangle $
                  & $\langle \tilde{v}\tilde{v}\rangle$
                  & $\langle \psi^\dagger \psi^2\rangle $ 
   \\
    \noalign{\smallskip}\hline\noalign{\smallskip}
    $d_\Gamma$ & $2$ & $1$
         & $2$ & $1$ & $2$
             & $0$ 
               \\
    $\delta_\Gamma$ & $2$ & $1$
         & $1$ & $0$ & $0$
             & $0$ 
              \\
    \noalign{\smallskip}\hline
  \end{tabular}
\end{table}
This theoretical analysis leads to the following renormalization of the parameters $g_0, D_0$ and $u_0$:
\begin{eqnarray}
& &  g_1 = g_{10}\mu^{-2\epsilon} Z_1^3,\quad g_2=g_{20}\mu^{2\Delta}Z_1^3 Z_3^{-1}, \nonumber\\*
& & u=u_0 Z_1 Z_2^{-1},\qquad\lambda = \lambda_0 \mu^{2\Delta} Z_2 Z_4^{-1}, \nonumber\\*
& & \nu=\nu_0 Z_1^{-1},\qquad\hskip0.35cm D=D_0Z_2^{-1},
 \label{eq:ren_rel}
\end{eqnarray}
where $\mu$ is the reference mass scale in the MS scheme~\cite{Zinn-Justin89}
and we have introduced the inverse Prandtl number $u=D/\nu$ for convenience.
From Table \ref{tab:canon_dims} it follows that $u$ is purely dimensionless quantity ($d^k_u=d^\omega_u=d_u=0$)
and physically it 
represents the ratio between diffusion and viscosity in a liquid.
In terms of the introduced renormalized parameters the total renormalized action for the annihilation reaction in a fluctuating
velocity field is
\begin{eqnarray}
 S_R &=& \int d{\mathbf x} dt \biggl\{
  \psi^\dagger\partial_t\psi +\psi^\dagger\nabla({\mathbf v}\psi)-u\nu Z_2\nabla^2\psi\nonumber\\*
  & + & \lambda u \nu \mu^{-2\Delta} Z_4[2\psi^\dagger+(\psi^\dagger)^2]\psi^2
  \nonumber\\*
  & - &\frac{1}{2} \tilde{{\mathbf v}}[g_1\nu^3\mu^{2\epsilon}(-\nabla^2)^{1-\Delta-\epsilon} -
  g_2\nu^3\mu^{-2\Delta}Z_3\nabla^2]\tilde{{\mathbf v}}\nonumber\\*
  & + & \tilde{{\mathbf v}} \cdot [\partial_t{\mathbf v}+({\mathbf v}\cdot\nabla){\mathbf v}-
  \nu Z_1\nabla^2{\mathbf v}] \biggl\}+n_0\int d{\mathbf x}    \psi^\dagger({\mathbf x},0).\nonumber\\
  \label{eq:total_act}
\end{eqnarray}
The renormalization constants $Z_i,i=1,2,3,4$ are to be calculated perturbatively through
the calculation of the UV divergent parts of the 1PI functions
$\Gamma_{\tilde{v}v}$, $\Gamma_{\tilde{v}\tilde{v}}$,
$\Gamma_{\psi^\dagger\psi}$, $ \Gamma_{\psi^\dagger\psi^2}$ and $\Gamma_{(\psi^\dagger)^2\psi^2}$. Interaction terms
corresponding to these functions have to be
added to the original action $S=S_1+S_2$ with the aim to ensure UV finiteness of all Green
functions generated by the renormalized action $S_R$. At this stage the main goal is to
calculate the renormalization constants $Z_i,i=1,2,3,4$.

The singularities in various Green functions will be realized in
the form of poles in $\epsilon$ and $\Delta$ and their linear combinations such as $2\epsilon+\Delta$ or
$\epsilon-\Delta$. Recall that for the consistency of the $\overline{\rm MS}$ scheme it is necessary that the
ratio
\[ \xi=\frac{\Delta}{\epsilon} \]
is a finite real number.
It should be noted that the graphs corresponding to
$\Gamma_{\psi^\dagger\psi^2}$ and $\Gamma_{(\psi^\dagger)^2\psi^2}$
differ only by one external vertex and thus give rise to equal renormalization of the rate constant $\lambda_0 D_0$.
Therefore, in what follows, we will always consider the function $\Gamma_{\psi^\dagger\psi^2}$.
In order to calculate the renormalization constants $Z_2$ and $Z_4$ we proceed according to the general scheme
suggested in \cite{Adzhemyan3}. We require the fulfillment of UV finiteness (i.e.finite limit
when $\epsilon,\Delta\rightarrow 0$ ) of the 1PI functions $\Gamma_{\psi^\dagger\psi}|_{\omega=0}$
and $\Gamma_{\psi^{\dagger}\psi^2}|_{\omega=0}$.

Because
the divergent part of the Feynman graphs should not depend on the value of $\omega$, we have adopted
the simplest choice $\omega=0$. It is convenient to introduce the dimensionless
expansion variables of the perturbation theory as
\begin{equation}
  \alpha_{10} \equiv \frac{g_{10}\overline{S_d}}{p^{2\epsilon}}\,,\quad
   \alpha_{20}\equiv\frac{g_{20}\overline{S_d}}{p^{-2\Delta}}\,,\quad
      \alpha_{30} \equiv \frac{\lambda_0\overline{S_d}}{p^{-2\Delta}},
  \label{eq:exp_par1}
\end{equation}
where $S_d$ is the surface area of the unit sphere in $d-$ dimensional space, $p$ is the total
momentum flowing into the Feynman diagram and $\overline{S_d}=S_d/(2\pi)^d$.
For brevity, in the following we use the abbreviation $\overline{g}_0\equiv g_0\overline{S_d}$ for
the parameters $\{g_{10},g_{20},\lambda_0 \}$ or their renormalized counterparts, respectively.

Next we present perturbation series for the 1PI Green functions to the
second order approximation.
The perturbative expansion for $\Gamma_{\psi^\dagger \psi}$ may be written as
\begin{equation}
 \Gamma_{\psi^{\dagger} \psi}|_{\omega=0} = D_0 p^2
 \biggl[-1+\sum_{\begin{subarray}{l}
                    n_1,n_2\ge 0,\\ n_1+n_2\ge 1
                 \end{subarray}}^{n_1+n_2=2}
 \alpha_{10}^{n_1}
\alpha_{20}^{n_2}\gamma_{\psi^{\dagger}\psi}^{(n_1,n_2)}(d,u_0) \biggl],
\label{eq:GF_prop}
\end{equation}
where $\gamma_{\psi^\dagger\psi}$ are dimensionless coefficients which contain poles in $\epsilon$ and $\Delta$. The explicit
dependence on the space dimension $d$ and the inverse Prandtl number $u_0$ is emphasized. It is important
to note that there are no terms in this series proportional to the expansion parameter $\alpha_{30}$. In terms
of the renormalized parameters the perturbative expansion for the Green function is
(\ref{eq:GF_prop})
\begin{equation}
   \frac{\Gamma_{\psi^{\dagger}\psi}|_{\omega=0}}{Dp^2} = Z_2 \biggl[-1+\sum_{\begin{subarray}{l}
                    n_1,n_2\ge 0,\\ n_1+n_2\ge 1
                 \end{subarray}}^{n_1+n_2=2}
   \alpha_{1}^{n_1} \alpha_{2}^{n_2}  \gamma_{\psi^{\dagger}\psi}^{(n_1,n_2)}(d,u) \biggl],
  \label{eq:GF_renprop}
\end{equation}
with the renormalized parameters $\alpha_{1}=  \overline{g_1} s^{2\epsilon} Z_1^{-3}$
and $\alpha_{2}= \overline{g_2} s^{-2\Delta} Z_3 Z_1^{-3}$ in
accordance with relations (\ref{eq:ren_rel}) and (\ref{eq:exp_par1}), where $s\equiv \mu/p$.
Here, we would like to stress, that in order to get the correct expansion in $\epsilon$ and $\Delta$, one
has to make replacement
\begin{equation}
d\rightarrow 2+2\Delta, \qquad u_0\rightarrow Z_1^{-1}Z_2 u,
\label{eq:change}
\end{equation}
 in the arguments of $\gamma_{\psi^{\dagger}\psi}^{(n_1,n_2)}$.
In the same way, the perturbation expansion series for the Green function $\Gamma_{\psi^\dagger \psi^2}$ is
\begin{eqnarray}
\Gamma_{\psi^\dagger \psi^2} |_{\omega =0}& = & -4 D_0 \lambda_0\biggl[1+\sum_{\begin{subarray}{l}
                    n_1,n_2,n_3\ge 0,\\ n_1+n_2+n_3\ge 1
                 \end{subarray}}^{n_1+n_2+n_3=2}
\alpha_{10}^{n_1}\alpha_{20}^{n_2} \alpha_{30}^{n_3} \nonumber\\*
& \times &\gamma_{\psi^\dagger \psi^2}^{(n_1,n_2,n_3)} (d,u_0) \biggl],
\label{eq:GF_ver}
\end{eqnarray}
where $\gamma_{\psi^\dagger \psi^2}$ are dimensionless coefficients resulting from calculation of the relevant Feynman graphs.
Again by replacing the bare parameters with the renormalized counterparts the following series is obtained
\begin{eqnarray}
 \frac{ \Gamma_{\psi^{\dagger} \psi^2} |_{\omega =0}}{4\lambda D \mu^{-2\Delta}} & = &
  -Z_4 \biggl[1+\sum_{\begin{subarray}{l}
                    n_1,n_2,n_3\ge 0,\\ n_1+n_2+n_3\ge 1
                 \end{subarray}}^{n_1+n_2+n_3=2}
   \alpha_{1}^{n_1} \alpha_{2}^{n_2} \alpha_{3}^{n_3}\nonumber\\*
   & \times & \gamma_{\psi^{\dagger}\psi^2}^{(n_1,n_2,n_3)} (d,u) \biggl],
\label{eq:GF_renver}
\end{eqnarray}
where the dimensionless parameter $\alpha_{3}= \overline{\lambda} s^{-2\Delta} Z_2^{-1} Z_4$ is introduced
and the change (\ref{eq:change}) is understood. Perturbation series for the Green
function $\Gamma_{(\psi^{\dagger})^2 \psi^2} $ has the same form, so
we do not present it here.

Denoting by $Z^{(n)}$ the contribution of the order $g^n,g=\{g_1,g_2,\lambda \}$, the first order
of renormalization constants $Z_2$ and $Z_4$ may be calculated via equations
\begin{eqnarray}
 Z_2^{(1)}  &=& \mathcal{L}[ \overline{g_1} s^{2\epsilon}\gamma_{\psi^{\dagger}\psi}^{(1,0)}
  +\overline{g_2} s^{-2\Delta}\gamma_{\psi^{\dagger}\psi}^{(0,1)} ],\\*
 Z_4^{(1)} &=& - \mathcal{L} [\overline{g_1}s^{2\epsilon}\gamma^{(1,0,0)}_{\psi^\dagger \psi^2}+
 \overline{g_2}s^{-2\Delta}\gamma^{(0,1,0)}_{\psi^\dagger \psi^2} \nonumber\\*
  & + & \overline{\lambda} s^{-2\Delta}
 \gamma^{(0,0,1)}_{\psi^\dagger \psi^2}],
 \label{eq:Z_ord1}
\end{eqnarray}
 where $\mathcal{L}$ stands for the operation of extraction of the UV-divergent part (poles in $\epsilon$ and $\Delta$ or
their linear combination).  In the $\overline{\rm MS}$ scheme  finite terms are discarded, so we do not need to take care of them.
At the second order the term for $Z_2$ can be schematically written as
\begin{eqnarray}
   Z_2^{(2)} & = &  \mathcal{L}\biggl[
  -\frac{ \overline{g_1} s^{2\epsilon} }{1+u}\biggl(u Z_2^{(1)}+(u+2)Z_1^{(1)}\biggl)
 \gamma_{\psi^{\dagger}\psi}^{(1,0)}  \nonumber\\*
& - & \overline{g_2} s^{-2\Delta} \biggl(\frac{u}{1+u} Z_2^{(1)}+\frac{u+2}{1+u}Z_1^{(1)} - Z_3^{(1)} \biggl)
\gamma_{\psi^{\dagger}\psi}^{(0,1)} \nonumber\\*
 & + & \overline{g_1}^2 s^{4\epsilon}
 \gamma_{\psi^{\dagger}\psi}^{(2,0)}+ \overline{g_1g_2} s^{2\epsilon-2\Delta}\gamma_{\psi^{\dagger}\psi}^{(1,1)}
 +\overline{g_2}^2 s^{-4\Delta}\gamma_{\psi^{\dagger}\psi}^{(0,2)}\biggl]. \nonumber\\*
\end{eqnarray}
\begin{figure}
  \resizebox{0.75\columnwidth}{!}{%
  \includegraphics{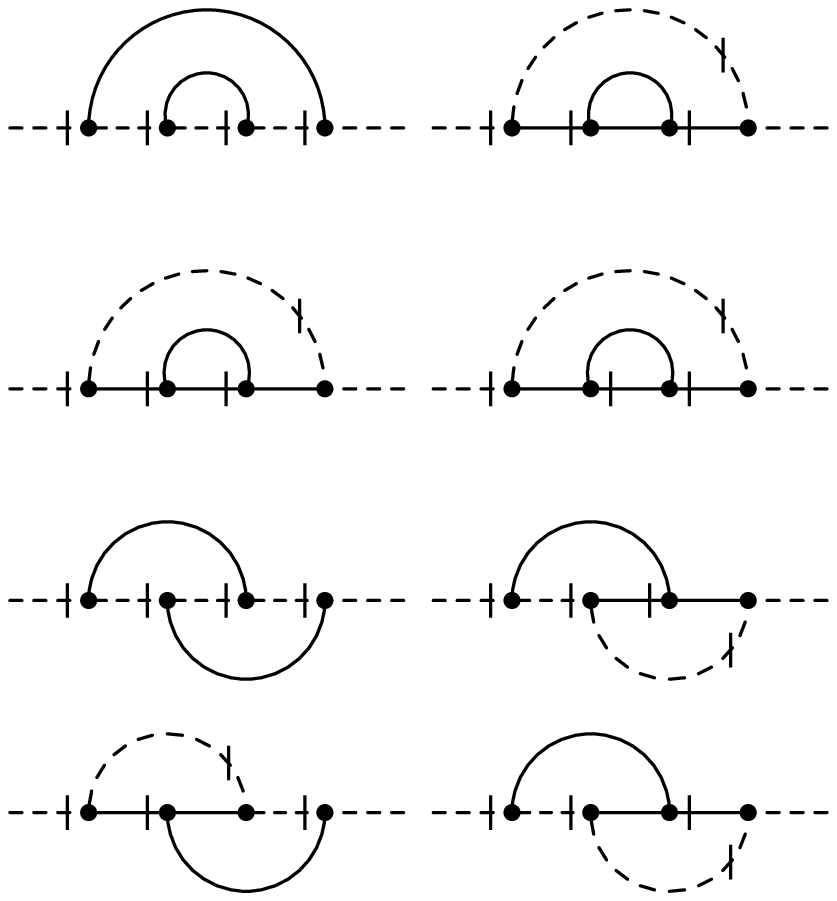}
  }
  \caption{Two-loop graphs for the per\-tur\-ba\-tion expan\-sion of 1PI function $\Gamma_{\psi^\dagger\psi}$.}
  \label{fig:loop_prop}
\end{figure}

The two-loop graphs that contribute to the calculation of $Z_2$ are represented by the
graphs depicted in Fig. \ref{fig:loop_prop}.
For the renormalization constants $Z_4$ we have the expression
\begin{eqnarray}
 Z_4^{(2)} &=& - \mathcal{L}[
   \overline{g_1}\overline{\lambda} s^{2(\epsilon-\Delta)} \gamma^{(1,0,1)}_{(\psi^{+})^2 \psi^2}
 +\overline{g_2}\overline{\lambda} s^{-4\Delta}\gamma^{(0,1,1)}_{\psi^\dagger \psi^2}
  \nonumber\\*
 & + & \overline{\lambda}^2 s^{-4\Delta}\gamma^{(0,0,2)}_{\psi^\dagger \psi^2} +
 \overline{g_1} s^{2\epsilon} \gamma^{(1,0,0)}_{\psi^\dagger \psi^2} (-3Z_1^{(1)})
 \nonumber\\*
 & + & \overline{g_2}s^{-2\Delta} \gamma^{(0,1,0)}_{\psi^\dagger \psi^2}(Z_3^{(1)}-3Z_1^{(1)})
 \nonumber\\*
 & + & \overline{\lambda}s^{-2\Delta}\gamma^{(0,0,1)}_{\psi^\dagger \psi^2}(2Z_4^{(1)}-Z_2^{(1)})].
 \label{eq:Z_ord2}
\end{eqnarray}

\begin{figure}
  \resizebox{0.75\columnwidth}{!}{%
  \includegraphics{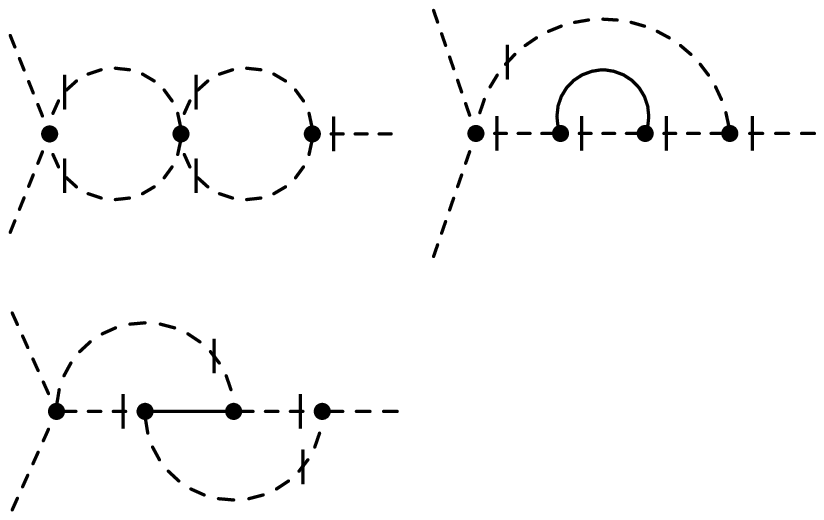}
  }
  \caption{Two-loop graphs for the perturbation expansion of 1PI function $\Gamma_{\psi^\dagger\psi^2}$.}
  \label{fig:loop_vert}       
\end{figure}

The two-loop graphs that contribute to the calculation of $Z_4$ are represented by the
graphs depicted in Fig. \ref{fig:loop_vert}.
From these expressions the renormalization constants $Z_2$ and $Z_4$ can be calculated in the form
\begin{eqnarray}
  Z_2 & = &  1  -\frac{\overline{g_1}}{8u(1+u)\epsilon}+\frac{\overline{g_2}}{8u(1+u)\Delta}
 +\frac{A_{11}\overline{g_1}^2}{\epsilon^2}
+ \frac{A_{22}\overline{g_2}^2}{\Delta^2}
\nonumber\\*
& &+\frac{A_{12}\overline{g_1g_2}}{\epsilon\Delta}
+\frac{B_{11}\overline{g_1}^2}{\epsilon}
+\frac{B_{22}\overline{g_2}^2}{\Delta}
+\frac{B_{12}\overline{g_1g_2}}{\epsilon-\Delta}
,
\label{eq:Z2}
\end{eqnarray}
\begin{eqnarray}
  Z_4 & = & 1 - \frac{\overline{\lambda}}{2\Delta} - \frac{1}{16u(1+u)}
\frac{\overline{g_1\lambda}}{(\epsilon-\Delta)\Delta} +
\frac{1}{32u(1+u)} \nonumber\\*
& \times  &\frac{\overline{g_2\lambda}}{\Delta^2}
  +\frac{\overline{\lambda}^2}{4\Delta^2}-\biggl(\frac{\overline{g_1\lambda}}{\epsilon-\Delta}-
   \frac{\overline{g_2\lambda}}{\Delta}\biggl)C(u,\xi).
   \label{eq:Z4}
\end{eqnarray}
Rather lengthy expressions for the coefficient functions $A_{ij}(\xi,u)$, $B_{ij}(\xi,u)$ and $C(u,\xi)$
can be found in Appendix \ref{app:ABcoeff}.

In a similar way we obtain renormalization constants $Z_1$ and $Z_3$ \cite{Adzhemyan3} from
condition of the UV finiteness
for the 1PI Green functions $\Gamma_{\tilde{v}v}|_{\omega=0}$
and $\Gamma_{\tilde{v}\tilde{v}}|_{\omega=0}$. The perturbation series for $\Gamma_{\tilde{v}v}$ can be written as
\begin{eqnarray}
  & & \Gamma_{\tilde{v}v}|_{\omega=0}=\nu_0 p^2 P_{ij}^p\biggl[-1+\sum_{\begin{subarray}{l}
                    n_1,n_2\ge 0,\\ n_1+n_2\ge 1
                 \end{subarray}}^{n_1+n_2=2}
 \alpha_{10}^{n_1}
\alpha_{20}^{n_2}\gamma_{\tilde{v}v}^{(n_1,n_2)}(d) \biggl], \nonumber\\*
\end{eqnarray}
and for $\Gamma_{\tilde{v}\tilde{v}}$ as
\begin{eqnarray}
  \Gamma_{\tilde{v}\tilde{v}}|_{\omega=0} & = & P_{ij}^p\biggl[g_{10}\nu_0^3p^{2-2\Delta-2\epsilon}+
 g_{20}\nu_0^3p^2 \biggl\{-1 \nonumber\\*
  & + & \sum_{\begin{subarray}{l} n_1\ge 0,n_2\ge -1,\\ n_1+n_2\ge 1
                 \end{subarray}}^{n_1+n_2=2}
 \alpha_{10}^{n_1}
\alpha_{20}^{n_2}\gamma_{\tilde{v}\tilde{v}}^{(n_1,n_2)}(d) \biggl\}
\biggl].
\end{eqnarray}
From the definition of the projection operator $P_{ij}^p$ it is easy to see that after
contracting indices $i$ and $j$ we are left with the constant $d-1$. Hence, rewriting
perturbation series for $\Gamma_{\tilde{v}v}$ and $\Gamma_{\tilde{v}\tilde{v}}$ in
the renormalized variables (\ref{eq:ren_rel}) and contracting indices $i$ and $j$ we obtain
\begin{eqnarray}
  & & \frac{\Gamma_{\tilde{v}v}|_{\omega=0}}{\nu p^2(d-1)}=
   -Z_1+Z_1\sum_{\begin{subarray}{l}
                    n_1,n_2\ge 0,\\ n_1+n_2\ge 1
                 \end{subarray}}^{n_1+n_2=2}
 \alpha_{1}^{n_1}
\alpha_{2}^{n_2}\gamma_{\tilde{v}v}^{(n_1,n_2)}(d) \biggl], \nonumber\\*
\end{eqnarray}
\begin{eqnarray}
  \frac{\Gamma_{\tilde{v}\tilde{v}}|_{\omega=0}}{(d-1)g_2\nu^3\mu^{-2\Delta}p^2}
& = & \frac{g_1}{g_2}s^{2\epsilon+2\Delta}+Z_3+Z_3  \nonumber\\*
  &\times & \sum_{\begin{subarray}{l} n_1\ge 0,n_2\ge -1,\\ n_1+n_2\ge 1
                 \end{subarray}}^{n_1+n_2=2}
 \alpha_{1}^{n_1}
\alpha_{2}^{n_2}\gamma_{\tilde{v}\tilde{v}}^{(n_1,n_2)}(d)
.\nonumber\\
\end{eqnarray}
Explicit expressions for the renormalization constants $Z_1$ and $Z_3$ are obtained
by the same algorithm as described above in detail for the calculation $Z_2$ and $Z_4$. Results for them
in the $\overline{\text{MS}}$ scheme can be found in \cite{Adzhemyan3}.
\section{IR stable fixed points and scaling regimes}
The coefficient functions of the RG differential operator for the Green functions
\begin{equation}
  D_{{\rm RG}} = \mu\frac{\partial}{\partial\mu}\biggl|_0 = \mu\frac{\partial}{\partial \mu} +\sum_{g_i}
 \beta_{i}\frac{\partial}{\partial g_i}
  -\gamma_1 \nu\frac{\partial}{\partial\nu}
  \label{eq:RGeq}
\end{equation}
are defined as
\begin{equation}
  \gamma_1=\mu\frac{\partial  \ln Z_1}{\partial\mu}\biggl|_0\,,\qquad
\beta_{i}=\mu\frac{\partial g_i}{\partial\mu}\biggl|_0\,,
\label{eq:RGfunctions}
\end{equation}
with the charges $g_i=\{g_1,g_2,u,\lambda \}$.
 In (\ref{eq:RGeq}) and (\ref{eq:RGfunctions}), the subscript
''0'' reminds that
partial derivatives are taken at fixed values of the bare parameters

From definitions (\ref{eq:RGfunctions})
  and the renormalization relations (\ref{eq:ren_rel}) it follows that
\begin{eqnarray}
& &  \beta_{g_1} = g_1(-2\epsilon+3\gamma_1)\,,\qquad \beta_{g_2} = g_2(2\Delta+3\gamma_1-\gamma_3), \nonumber\\*
& & \beta_\lambda = \lambda(2\Delta-\gamma_4+\gamma_2)\,,\quad \beta_u = u(\gamma_1-\gamma_2),
   \label{eq:exp_beta}
\end{eqnarray}
where the anomalous dimensions $\gamma_\alpha$ ($\alpha=2,3,4$) are defined
similarly as $\gamma_1$ as
\begin{equation}
    \gamma_\alpha=\mu\frac{\partial  \ln Z_\alpha}{\partial\mu}\biggl|_0.
    \label{eq:def_gamma}
\end{equation}
We are interested in the IR asymptotics of small momentum ${\bf p}$ and frequencies $\omega$
of the renormalized functions or, equivalently, large relative distances and time differences in
the $(t,{\bf x})$ representation. Such a behavior is governed by the IR-stable fixed point $g^*=(g_1^*,g_2^*,u^*,\lambda^*)$, which are determined as zeroes of the $\beta$ functions
$\beta(g^*)=0$. The fixed point $g^*$ is IR stable, if real parts of all eigenvalues
of the matrix $\omega_{ij}\equiv \partial \beta_i/\partial g_j|_{g=g^*}$
are strictly positive.

From the
explicit form 
 of the renormalization constants
$Z_2$ and $Z_4$ (\ref{eq:Z2}-\ref{eq:Z4}) and definitions (\ref{eq:RGfunctions}-\ref{eq:def_gamma})
 it is possible to calculate anomalous dimensions
$\gamma_2$ and $\gamma_4$
\begin{eqnarray}
  \label{eq:gamma2}
     \gamma_2 & = &   \frac{\overline{g_1}+\overline{g_2}}{4u(1+u)}  -4B_{11}\overline{g_1}^2+
4B_{22}\overline{g_2}^2 - 2 B_{12}\overline{g_1g_2},\\
  \gamma_4 & = &  -\overline{\lambda} + \overline{\lambda}(\overline{g_1}+\overline{g_2})C(u,\xi).
  \label{eq:gamma4}
\end{eqnarray}
A straightforward calculation shows that higher order poles cancel each other, so
 that the anomalous dimensions $\gamma_2$ and $\gamma_4$ are finite.
  For completeness we quote also anomalous dimensions $\gamma_1$ and
$\gamma_3$ \cite{Adzhemyan3} to the same order
\begin{eqnarray}
     \gamma_1 & = &  \frac{\overline{g_1}+\overline{g_2}}{16} + \frac{(4\xi+3)}{512(2+\xi)}\overline{g_1}^2 +
  \frac{5\xi+3}{512}\overline{g_1g_2} \nonumber\\*
  & - & \frac{R}{256}(\overline{g_1}+\overline{g_2})^2,
  \label{eq:gamma1}
\end{eqnarray}
\begin{eqnarray}
  \gamma_3 & = &  \frac{(\overline{g_1}+\overline{g_2})^2}{16\overline{g_2}} -
  \frac{\xi(13+19\xi)}{1024(2+\xi)} \frac{\overline{g_1}^3}{\overline{g_2}} +
  \frac{34\xi+19+6\xi^2}{512(2+\xi)}\overline{g_1}^2 \nonumber\\*
  & - & \frac{3\overline{g_1}^2}{512}+
  \frac{13+31\xi}{1024}\overline{g_1g_2} +
 \frac{1-R}{256}\frac{(\overline{g_1}+\overline{g_2})^3}{\overline{g_2}},
  \label{eq:gamma3}
\end{eqnarray}
where the value $R=-0.168$ is a result from numerical integration.
Zeroes of the beta functions (\ref{eq:exp_beta}) determine possible IR behavior of the model.
There are four IR stable fixed points and
 one IR unstable fixed point. In this section we present them with their regions of stability.

({\it i}$\,$) The trivial (Gaussian) fixed point
\begin{equation}
\label{Gaussian}
\overline{g_1}^*=\overline{g_2}^*=\overline{\lambda}^*=0\,,
\end{equation}
with no restrictions on the inverse Prandtl number $u$. The Gaussian fixed point is stable, when
\begin{equation}
\label{Gstable}
\epsilon<0\,,\qquad \Delta >0\,.
\end{equation}
and physically corresponds to the case, when the mean-field solution is valid and fluctuation
effects negligible.
({\it ii}$\,$)
The short-range (thermal) fixed point
\begin{eqnarray}
\label{SR}
& & \overline{g_1}^*=0\,,\quad \overline{g_2}^*=-16\Delta+8(1+2R)\Delta^2,\nonumber\\*
& & u^*=\frac{\sqrt{17}-1}{2}-1.12146\Delta,\nonumber\\*
& &\overline{\lambda}^*=-\Delta+\frac{\Delta^2}{2}\,(\xi-2.64375),
\end{eqnarray}
at which local correlations of the random force dominate over the long-range
correlations. This fixed point
has the following basin of attraction
\begin{eqnarray}
\label{SRstable}
& & \Delta-\frac{2R-1}{2}\Delta^2<0,\quad 2\epsilon+3\Delta-\frac{3\Delta^2}{2} <0,\\*
& & \Delta+\frac{1}{2}\Delta^2<0,\quad \Delta+0.4529\Delta\epsilon<0
\end{eqnarray}
and corresponds to anomalous decay faster than that due to density fluctuations only, but
slower than the mean-field decay.

({\it iii}$\,$)
The kinetic~\cite{Hnatich99} fixed point with finite rate coefficient:
 \begin{eqnarray}
& & \overline{g_{1}}^{\ast}=
 \frac{32}{9}\,\frac{\epsilon\,(2\epsilon+3\Delta)}
 {\epsilon+\Delta}+g_{12}^*(\xi)\epsilon^2,\nonumber\\*
& &  \overline{g_{2}}^{\ast}=\frac{32}{9}\,
 \frac{\epsilon^2}{\Delta+\epsilon}+g_{22}^*(\xi)\epsilon^2,\nonumber\\*
& & u^*=\frac{\sqrt{17}-1}{2}+u_1^*(\xi)\epsilon, \nonumber\\*
& & \overline{\lambda}^*=-\frac{2}{3}(\epsilon+3\Delta)+\frac{1}{9\pi}(3\Delta+\epsilon)(Q\epsilon-\Delta),
\label{K1}
 \end{eqnarray}
 Explicit expressions for functions $g_{12}^*(\xi)$, $g_{22}^*(\xi)$ and $u_1^*(\xi)$ are given in
 Appendix \ref{app:fixedpoints}. In (\ref{K1}) the constant
 $Q=1.64375$. The fixed point (\ref{K1}) is stable, when inequalities
\begin{equation}
\label{K1stable}
\text{\rm{Re
}}
\Omega_{\pm}>0\,,\qquad\epsilon>0\,,\qquad -\frac{2}{3}\epsilon<\Delta<-\frac{1}{3}\epsilon,
\end{equation}
are fulfilled, where
\begin{eqnarray}
  \Omega_{\pm} & = & \Delta+\frac{4}{3}\epsilon\pm\frac{\sqrt{9\Delta^2-12\epsilon\Delta-8\epsilon^2}}{3} +
\frac{2}{9}\biggl(-(3+2R)\epsilon^2\nonumber\\*
& & - 3\epsilon\Delta+
    \frac{4\epsilon(\epsilon+3\Delta)R-6\epsilon^2-12\epsilon\Delta-9\Delta^2}{\sqrt{9\Delta^2-12\Delta-8\epsilon^2}}\epsilon \biggl)
\end{eqnarray}
The decay rate controlled by this fixed point
of the average number density is faster than the decay rate induced by dominant local
force correlations, but still slower than the mean-field decay rate.

({\it iv}$\,$)
The kinetic fixed point with vanishing rate coefficient:
 \begin{eqnarray}
 & & \overline{g_{1}}^{\ast}=
 \frac{32}{9}\,\frac{\epsilon\,(2\epsilon+3\Delta)}
 {\epsilon+\Delta}+g_{12}^*(\xi)\epsilon^2,\nonumber\\*
& &  \overline{g_{2}}^{\ast}=\frac{32}{9}\,
 \frac{\epsilon^2}{\Delta+\epsilon}+g_{22}^*(\xi)\epsilon^2, \nonumber\\*
& & u^*=\frac{\sqrt{17}-1}{2}+u_1^*(\xi)\epsilon\,,\quad  \overline{\lambda}^*=0\,.
\label{K2}
 \end{eqnarray}
This fixed point is stable, when the long-range correlations of the random
force are dominant
\begin{equation}
\label{K2stable}
\mathrm{Re }
\Omega_{\pm}>0,\quad\epsilon>0,\quad \Delta>-\frac{1}{3}\epsilon\,,
\end{equation}
and corresponds to reaction kinetics with the normal (mean-field like) decay rate.

({\it v}$\,$)
Driftless fixed point given by
\begin{equation}
   \overline{g_1}^*=\overline{g_2}^*=0,\quad u^* \mbox{ not fixed},\quad \overline{\lambda}^*=-2\Delta,
\end{equation}
with the following eigenvalues
\begin{equation}
  \Omega_1=-2\epsilon,\quad \Omega_2=-\Omega_4=2\Delta,\quad \Omega_3=0.
\end{equation}
An analysis of
the structure of the fixed points and the basins of attraction leads to the following
physical picture of the effect of the random stirring on the
reaction kinetics. Anomalous behavior always emerges below two dimensions, when the
{\it local} correlations are dominant in the spectrum of the random forcing
[the short-range fixed point ({\it ii}$\,$)]. However, the random stirring
gives rise to an effective reaction rate faster than the density-fluctuation
induced reaction rate even in this case. The anomaly is present
(but with still faster decay, see Section \ref{sec:Long-time}) also, when the long-range part
of the forcing spectrum is effective, but the powerlike falloff of the correlations is
fast [this regime is governed by the kinetic fixed point ({\it iii}$\,$)].
Note that this is different from the case in which the divergenceless
random velocity field is time-independent, in which case there is no fixed
point with $\lambda^*\ne 0$\cite{Deem2}.
At slower spatial falloff of correlations, however, the anomalous reaction kinetics
is replaced by a mean-field-like behavior
[this corresponds to the kinetic fixed point ({\it iv}$\,$)]. In particular,
in dimensions $d>1$ this is the situation
for the value $\epsilon=2$ which corresponds to the Kolmogorov spectrum of the velocity field in
fully developed turbulence.
Thus, long-range correlated forcing gives rise to a random velocity field, which
tends to suppress the effect of density fluctuations on the reaction kinetics below two dimensions.

\begin{figure}
  \resizebox{0.75\columnwidth}{!}{%
  \includegraphics{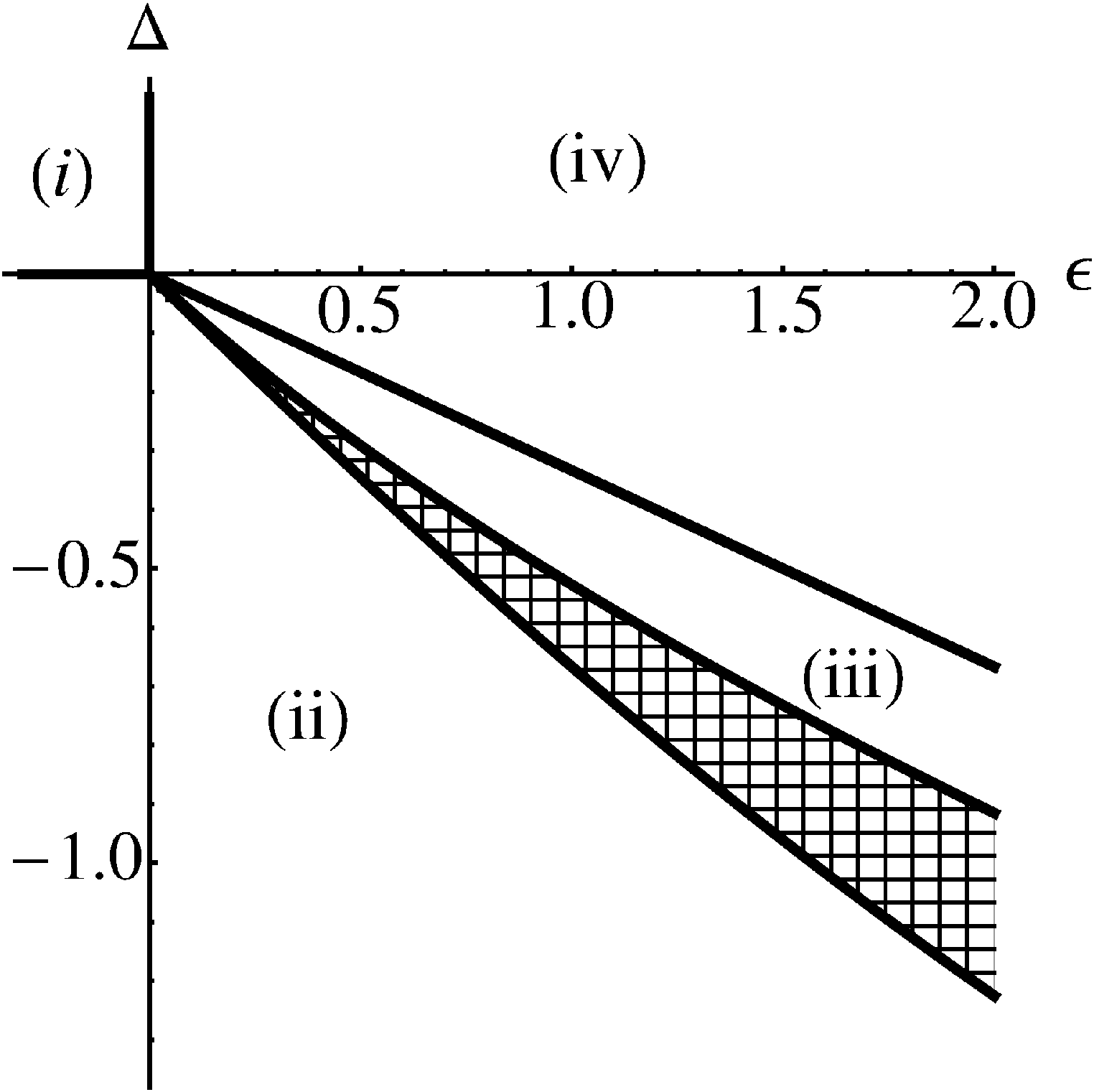}
  }
  \caption{Regions of stability for the IR stable fixed points $({\it i})-({\it iv})$ in $(\epsilon,\Delta)$ plane.}
  \label{fig:regions_of_stability}
\end{figure}

For better illustration, regions of stability for fixed points $({\it i})-({\it iv})$ are
depicted in Fig.\ref{fig:regions_of_stability}. Wee see
that in contrast to the one-loop approximation \cite{Hnatic}, overlap (dashed region)
between regions of stability of fixed points $({\it ii})$ and $({\it iii})$ is observed.
It is a common situation in the perturbative RG approach that higher order terms
lead to either gap or overlap between neighbouring stability regions. The physical realization
of the large-scale behavior then depends on the initial state of the system.
\section{Long-time asymptotics of number density}
\label{sec:Long-time}
Because the renormalization and calculation of the fixed points of
the RG are carried out at two-loop level, we are able to find the
first two terms of the $\epsilon$, $\Delta$ expansion of the average
number density, which corresponds to solving the stationarity
equations at the one-loop level. The simplest way to find the
average number density is to calculate it from the stationarity
condition of the functional Legendre transform~\cite{Vasiliev}
(which is often called the effective action) of the generating
functional
obtained by replacing the unrenormalized action by the renormalized
one in the weight functional. This is a convenient way to avoid any
summing procedures used~\cite{Lee94} to take into account the
higher-order terms in the initial number density $n_0$.

\subsection{Stationarity equations of the effective action}
We are
interested in the solution for the number density, therefore we put
the expectation values of the fields ${\mathbf v}$ and $\tilde{
{\mathbf v} }$ equal to zero at the outset (but retain, of course,
the propagator and the correlation function). Therefore,
at the second-order approximation 
the effective renormalized action for this model is
\begin{eqnarray}
\label{GammaAA1loop}
\Gamma_R & = &  S_1+\frac{1}{4}
\raisebox{-3.4ex}{ \epsfysize=1.25truecm
\epsffile{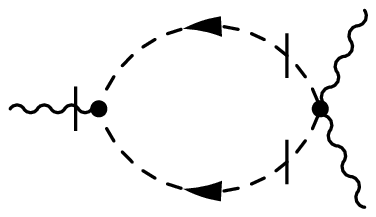}}
+\frac{1}{8}
\raisebox{-3.4ex}
{ \epsfysize=1.25truecm
\epsffile{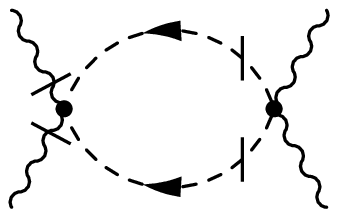}}
+
\nonumber\\*
& &\raisebox{-3.4ex}{ \epsfysize=1.25truecm
\epsffile{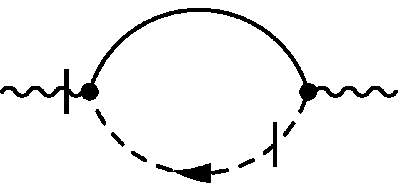}} +\ldots\,,
\end{eqnarray}
where $S_1$ is the action (\ref{eq:S_1}) (within our convention $S_2=0$ in the effective action)
and graphs are shown together with their symmetry coefficients.
The slashed wavy line corresponds to the field $\psi^\dagger$ and the single wavy line
to the field $\psi$.
 The stationarity equations for the variational functional
\begin{equation}
  \frac{\delta \Gamma_R}{\delta \psi^\dagger} = \frac{\delta \Gamma_R}{\delta \psi} = 0
  \label{eq:var_func}
\end{equation}
give
rise to the equations
\begin{multline}
\label{EqAA1}
\partial_t \psi = u\nu Z_2 \nabla^2 \psi-2\lambda u\nu \mu^{-2\Delta}Z_4 \left(1+\psi^\dagger\right)\psi^2\\
+4u^2\nu^2\lambda^2\mu^{-4\Delta}\int\limits_0^\infty\!dt' \int d\my \,( \Delta^{\psi\psi^\dagger})^2(t-t',\mx-\my)
 \psi^2(t',\my)\\
+4u^2\nu^2\lambda^2\mu^{-4\Delta}\psi^\dagger(t,\mx)\\*
\times\int\limits_0^\infty\!dt'\int\!d \my \,( \Delta^{\psi\psi^\dagger})^2(t-t',\mx-\my)\psi^2(t',\my)\\
+\doo{}{x_i}\int\limits_0^\infty\!dt' \int \my \, \Delta_{ij}^{vv}(t-t',\mx-\my)\\*
\times\doo{}{x_j}\Delta^{\psi\psi^\dagger}(t-t',\mx-\my) \psi(t',\my)
+\ldots,
\end{multline}
\begin{multline}
-\partial_t \psi^\dagger= u\nu Z_2\nabla^2\psi^\dagger-2\lambda u\nu\mu^{-2\Delta}Z_4\left[2\psi^\dagger +\left(\psi^\dagger\right)^2\right]\psi
\\*
+8u^2\nu^2\lambda^2\mu^{-4\Delta}\int\limits_0^\infty\!dt'\int \!d \my( \Delta^{\psi\psi^\dagger})^2(t'-t,\my-\mx)
\psi^\dagger (t',\my)\times
\\*
\psi(t,\mx)+4u^2\nu^2\lambda^2\mu^{-4\Delta}\int\limits_0^\infty\!dt'\int\!d \my( \Delta^{\psi\psi^\dagger})^2(t'-t,\my-\mx)\\
\times\left[\psi^\dagger (t',\my)\right]^2 \psi(t,\mx)\\*
+\int\limits_0^\infty\!dt' \int \!d \my \, \Delta_{ji}^{vv}(t'-t,\my-\mx)\\*
\times\doo{}{x_i}\Delta^{\psi\psi^\dagger}(t'-t,\my-\mx) \doo{}{y_j}\psi^\dagger(t',\my)
+\ldots
\label{EqAA2}
\end{multline}
In (\ref{EqAA1}) and (\ref{EqAA2}), in the integral terms it is sufficient to put all renormalization constants equal to unity.
Substituting the solution $\psi^\dagger=0$ of (\ref{EqAA2}) into (\ref{EqAA1}) we arrive at the
fluctuation-amended rate equation in the form
\begin{multline}
\label{RateEqAA1loop}
\partial_t \psi = u\nu Z_2 \nabla^2 \psi-2\lambda u\nu \mu^{-2\Delta}Z_4 \psi^2+\\*
4u^2\nu^2\mu^{-4\Delta}\lambda^2\int\limits_0^\infty\!dt' \int \!d \my \,( \Delta^{\psi\psi^\dagger})^2(t-t',\mx-\my) \psi^2(t',\my)\\
+\doo{}{x_i}\int\limits_0^\infty\!dt' \int \!d \my \, \Delta_{ij}^{vv}(t-t',\mx-\my)\\*
\times
\doo{}{x_j}\Delta^{\psi\psi^\dagger}(t-t',\mx-\my) \psi(t',\my)
+\ldots,
\end{multline}
This is a nonlinear partial integro-differential equation, whose explicit solution is not known.
It is readily seen that for a homogeneous solution the term resulting from the third graph in
(\ref{GammaAA1loop}) vanishes and
hence the influence of the velocity field on the homogeneous annihilation process would be only through the
renormalization of the coefficients $\lambda$ and $D$. However, in case of a nonuniform density field $\psi$
the effect of velocity fluctuations is explicit in (\ref{RateEqAA1loop}). Such a solution can be most probably
found only numerically.

To arrive at an analytic solution, we restrict ourselves to the homogeneous number density $n(t)=\langle\psi(t)\rangle$, which
can be identified with the expression (\ref{eq:aver_number}).
In this case the last term in (\ref{RateEqAA1loop}) vanishes together with the Laplace operator term and the remaining coordinate integral
may be calculated explicitly .
The propagator is the diffusion kernel of the renormalized model (we consider first the system in the general space dimension $d$)
\begin{multline}
 \Delta^{\psi\psi^\dagger}(t-t',\mx)\\*
 =\frac{\theta(t-t')}{\left[4\pi u\nu(t-t')\right]^{d/2}}\exp\left[-\frac{x^2}{4u\nu(t-t')}\right].
\label{eq:diffkernel}
\end{multline}
As noted above, for calculation of the one-loop contribution it is sufficient to put the renormalization constant $Z_2=1$ in the propagator
$\Delta^{\psi\psi^\dagger}$.
Therefore,
evaluation of the Gaussian coordinate integral in (\ref{RateEqAA1loop}) yields
\begin{equation}
\label{integral}
\int\!d \my\,( \Delta^{\psi\psi^\dagger})^2(t-t',\mx-\my)=\frac{\theta(t-t')}{\left[8\pi u\nu(t-t')\right]^{d/2}}
\end{equation}
and we arrive at the ordinary integro-differential equation
\begin{multline}
\label{RateEqAA1loopUniform}
\dee{ n(t)}{t}= -2\lambda u\nu \mu^{-2\Delta}Z_4  n^2(t)\\*
+
4\lambda^2u^2\nu^2\mu^{-4\Delta}\int\limits_0^t\!dt'\,\frac{n^2(t')}{\left[8\pi u\nu(t-t')\right]^{d/2}}\,.
\end{multline}
Spatial fluctuations in the number density show in the integral term and affect rather heavily
even the homogeneous solution. In particular, the integral in (\ref{RateEqAA1loopUniform}) diverges at
the upper limit in space dimensions $d\ge 2$. This is a consequence of the UV divergences in the model
above the critical dimension $d_c=2$ and near the critical dimension is remedied by the UV renormalization of the
model. To see this, subtract and add the term $n^2(t)$ in the integrand to obtain
\begin{multline}
\label{RateEqAA1loopUniform2}
\dee{ n(t)}{t}= -2\lambda u\nu \mu^{-2\Delta}Z_4  n^2(t)\\*
+
4\lambda^2u^2\nu^2\mu^{-4\Delta}n^2(t)\int\limits_0^t\!\frac{dt'}{\left[8\pi u\nu(t-t')\right]^{d/2}}\\*
+
4\lambda^2u^2\nu^2\mu^{-4\Delta}\int\limits_0^t\!dt'\,\frac{n^2(t')-n^2(t)}{\left[8\pi u\nu(t-t')\right]^{d/2}}\,.
\end{multline}
The last integral here is now convergent at least near two dimensions, provided the solution $n(t)$ is
a continuous function. This is definitely the case for the iterative solution constructed below.
The divergence in the first integral in (\ref{RateEqAA1loopUniform2}) may be explicitly calculated
below two dimensions and
is canceled -- in the leading order in the parameter $\Delta=(d-2)/2$ -- by the one-loop term of the renormalization constant $Z_4$
(\ref{eq:Z4}). Expanding the right-hand side of (\ref{RateEqAA1loopUniform2}) in the parameter $\Delta=(d-2)/2$ to the next-to-leading
order we arrive at the equation
\begin{multline}
\label{RateEqAA1loopUniform3}
\dee{ n(t)}{t}= -2\lambda  u\nu\mu^{-2\Delta}   n^2(t)\\*
+2\lambda u\nu\mu^{-2\Delta}    n^2(t)\left\{\frac{\lambda}{4\pi}\,\left[\gamma+\ln\left(2u\nu\mu^2 t\right)\right]\right\}\\*
+
\frac{\lambda^2u\nu\mu^{-2\Delta}}{2\pi}\int\limits_0^t\!dt'\,\frac{n^2(t')-n^2(t)}{t-t'}\,.
\end{multline}
without divergences near two dimensions. Here, the factor $\mu^{-2\Delta}$ has been retained intact in order not
to spoil the consistency of scaling dimensions in different terms of the equation. In (\ref{RateEqAA1loopUniform3}), $\gamma=0.57721$ is Euler's constant and
we have considered the coupling constant $\lambda$ and the parameter
$\Delta=(d-2)/2$ to be small parameters of the same order taking into account the magnitudes of the parameters in the basins of attraction
of the fixed points of the RG.

The leading-order approximation for $n(t)$ is given by the first term on the
right-hand side of (\ref{RateEqAA1loopUniform3})
in the form
\begin{equation}
\label{eq:LOSolution}
n^{(1)}(t)=
\frac{n_0}{1+2{\lambda} u\nu t\mu^{-2\Delta}n_0}\,,
\end{equation}
where $n_0$ is the initial number density.
After substitution of this expression the integral
term in (\ref{RateEqAA1loopUniform3}) is of the order of $\lambda^3$ and thus negligible in the present next-to-leading-order calculation.
Indeed, integration by parts in the integral in (\ref{RateEqAA1loopUniform3}) yields
\begin{multline}
\label{eq:integral}
\int\limits_0^t\!dt'\,\frac{n^2(t')-n^2(t)}{t-t'}=\lim_{\varepsilon\to 0}\int\limits_0^{t-\varepsilon}\!dt'\,\frac{n^2(t')-n^2(t)}{t-t'}\\
=
\left[n_0^2-n^2(t)\right]\ln t+2\int\limits_0^t\!dt'\,\ln(t-t')n(t')\dee{ n(t')}{t'}\,.
\end{multline}
From (\ref{RateEqAA1loopUniform3}) it is readily seen that $\displaystyle\dee{ n(t)}{t}=O\left(\lambda\right)$. From the explicit expression
(\ref{eq:LOSolution}) it follows that $n^2(t)-n_0^2=O\left(\lambda\right)$ as well. Therefore, in the iterative solution of
(\ref{RateEqAA1loopUniform3}) the term with the integral on the right side is of the order of $\lambda^3$.

In this
approximation, Eq. (\ref{RateEqAA1loopUniform3}) yields
\begin{multline}
\label{NLsolution}
n(t)=\\
\frac{n_0}{1+2{\lambda} u\nu t\left\{1+ \frac{\displaystyle\lambda}{\displaystyle 4\pi}\left[1-\gamma
-\ln\left(2u\nu\mu^2 t\right)\right]\right\}\mu^{-2\Delta}n_0}\,,
\end{multline}
where $n_0$ is the initial number density.

\subsection{Asymptotic behavior of the number density}
We shall use the next-to-leading order iterative solution (\ref{NLsolution}) of the stationarity equation
(\ref{RateEqAA1loopUniform3}) as the initial condition of the RG equation for the number density
to evaluate the long-time asymptotic behavior of  the number density. The RG equation is obtained by applying the
RG differential operator (\ref{eq:RGeq}) to the renormalized number density $n(t)$.

Since the fields $\Phi=\{v,\tilde{v},\psi,\psi^\dagger\}$ are not renormalized, the renormalized connected
Green functions $W$ differ from the unrenormalized $W_0=\langle\Phi\ldots\Phi\rangle$ \cite{Vasiliev04} only by the
choice of parameters and thus one may write
\begin{equation}
  W( g,\nu,\mu,\ldots) = W_0(g_0,\nu_0,\ldots),
  \label{eq:diff_GF}
\end{equation}
where $g_0=\{g_{10},g_{20},u_0,\lambda_0\}$ is the full set of the bare parameters and dots denotes
all variables unaffected by the renormalization procedure. The independence
of the renormalized number density $n(t)$
of the renormalization mass
parameter $\mu$ is expressed by the equation
$$D_{RG}n(t,\mu,\nu,n_0,g)= 0\,.$$
Using this equation
together with the explicit expression for the RG
differential operator (\ref{eq:RGeq})
the RG equation for the mean number density $n(t)$ is readily obtained:
\begin{equation}
\left(\mu\frac{\partial}{\partial\mu}+\sum\limits_{g_i}\beta_{g_i}\frac{\partial}{\partial g_i}
-\gamma_1\nu\frac{\partial}{\partial\nu}\right)n(t,\mu,\nu,n_0,g)=0.
 \label{eq:RG}
\end{equation}
We are interested in long-time behavior of the system ($t\rightarrow\infty$), therefore
we trade the
renormalization mass for the time variable.
According to the canonical dimensions of parameters and fields in Table \ref{tab:canon_dims},
the Euler equations with respect to the wave number and time assume the form
\cite{Adzhemyan99}
\begin{equation}
  \left(\mu\frac{\partial}{\partial \mu}-2\nu\frac{\partial}{\partial \nu}+
   dn_0\frac{\partial}{\partial n_0}-d \right) n(t,\mu,\nu,n_0,g)=0,
   \label{eq:scale_mom}
\end{equation}
\begin{equation}
  \left( -t\frac{\partial }{\partial t}+\nu\frac{\partial}{\partial \nu}\right)   n(t,\mu,\nu,n_0,g)=0,
   \label{eq:scale_freq}
\end{equation}
where the first equation expresses scale invariance with respect to wave number and the second equation
with respect to time.
Eliminating partial derivatives with respect to the renormalization mass $\mu$ and viscosity $\nu$
from (\ref{eq:RG}), (\ref{eq:scale_mom}) and (\ref{eq:scale_freq})
we obtain the Callan-Symanzik equation for the mean number density:
\begin{equation}
  \biggl[(2-\gamma_1)t\frac{\partial}{\partial t}+\sum_{g} \beta_g \frac{\partial}{\partial g}-
 dn_0 \frac{\partial}{\partial n_0}+d \biggr] n\left(t,\mu,\nu,n_0,g\right)=0
  \label{eq:Callan}
\end{equation}
To separate information given by the RG, consider the dimensionless normalized mean number density
\begin{equation}
\label{normaln}
\frac{n}{n_0}=\phi\left(\nu\mu^2 t,\lambda u\,\frac{n_0}{\mu^d},g\right)\,.
\end{equation}
For the asymptotic analysis, it is convenient to express the
initial
number density
$n_0$ as an argument of the function $\phi$
in the combination used here.

Solution of (\ref{eq:Callan}) by the method of characteristics yields
\begin{equation}
\phi\left(\nu\mu^2 t,\lambda u\,{n_0\over \mu^d},g\right)=\phi\left({\nu}\mu^2 \tau,\overline{\lambda}\overline{u}\,{\overline{n}_0\over \mu^d},\overline{g}\right)
 \label{eq:solNLRG}
\end{equation}
where $\tau$ is the time scale.
In Eq. (\ref{eq:solNLRG}), $\overline{g}$ and $\overline{n}_0$
are the first integrals of the system of differential equations
\begin{equation}
  t\frac{d}{dt}\overline{g}=-\frac{\beta_g (\overline{g})}{2-\gamma_1(\overline{g})}\,,\quad
  t\frac{d}{dt}\overline{n}_0=d\frac{\overline{n}_0}{2-\gamma_1(\overline{g})}.
  \label{eq:first_int}
\end{equation}
Here $\overline{g}=\{\overline{g_1},\overline{g_2},\overline{u},\overline{\lambda} \}$ with
initial conditions $\overline{g}|_{t=\tau}=g$ and $\overline{n}_0|_{t=\tau}=n_0$.
In particular,
\begin{equation}
\label{eq:initialdensity}
\overline{\lambda}\overline{u}\overline{n}_0=\lambda u\,n_0\,\left({t\over \tau}\right)\exp\left[{\displaystyle  \int_\tau^t{\displaystyle \gamma_4 ds\over\displaystyle  (2-\gamma_1)s}}\right]\,.
\end{equation}
From here it follows that
the asymptotic expression of the integral on the right-hand side of (\ref{eq:initialdensity}) in the
vicinity of the IR-stable fixed point $g^*$ is of the form
\begin{multline}
\label{eq:asyamplitude}
{ \int_\tau^t{\displaystyle \gamma_4 ds\over\displaystyle  (2-\gamma_1)s}}
\operatornamewithlimits{\sim}_{t\to\infty}{\gamma_4^*\over 2-\gamma_1^*}\ln\left({t\over\tau}\right)\\*+
{2\over 2-\gamma_1^*}\int\limits_\tau^\infty\!{ (\gamma_4-\gamma_4^*)ds\over  (2-\gamma_1)s}
={\gamma_4^*\over 2-\gamma_1^*}\,\ln\left({t\over \tau}\right)+\tilde{c}_4(\tau)\,,
\end{multline}
corrections to which vanish in the limit $t\to\infty$. In (\ref{eq:asyamplitude}) and henceforth, the
notation $\gamma_1^*=\gamma_1\left(g^*\right)$ has been used.

From the point of view of the long-time asymptotic
behavior the next-to-leading term in (\ref{eq:asyamplitude}) is an inessential constant.
From (\ref{eq:initialdensity}) and (\ref{eq:asyamplitude})
it follows that
in the vicinity of the fixed point
\begin{multline}
\label{eq:initialdensityasy}
\overline{\lambda}\overline{u}\,{\overline{n}_0\over \mu^d}
\sim \lambda u\,{n_0\over \mu^d} \left({t\over \tau}\right)^{\displaystyle 1+{\displaystyle \gamma_4^*\over \displaystyle 2-\gamma_1^*}}\widetilde{C}_n\\*
\equiv \lambda u\,{n_0\over \mu^d} \left({t\over \tau}\right)^\alpha \widetilde{C}_n
\equiv \overline{y}\,\widetilde{C}_n\,,
\end{multline}
where a shorthand notation $\overline{y}$ has been introduced for the long-time scaling of the normalized number density as well as
the dimensional normalization constant
\[
\widetilde{C}_n=e^{d\tilde{c}_4(\tau)}\,.
\]
and the decay exponent
\begin{equation}
\label{eq:defalpha}
\alpha=\displaystyle 1+{\displaystyle \gamma_4^*\over \displaystyle 2-\gamma_1^*}
\end{equation}
The asymptotic behavior of the normalized number density is described by the scaling function $f(x,y)$:
\begin{multline}
\phi\left(\nu\mu^2 t,\lambda u\,{n_0\over \mu^d},g\right)\\*
\sim\phi\left({\nu}\mu^2 \tau,\widetilde{C}_n\overline{y},{g}^*\right)\equiv
f\left({\nu}\mu^2 \tau,\widetilde{C}_n\overline{y}\right)\,.
\label{eq:scalingf}
\end{multline}
The scaling function $f(x,y)$ describing the asymptotic behavior of the normalized number density
$\phi=n/n_0$ is a function of two dimensionless argument only,
whereas the generic $\phi$ has six dimensionless arguments (all four coupling constants on top of the
scaling arguments of $f(x,y)$).
We recall that the generic solution of the Callan-Symanzik equation (\ref{eq:Callan}) does not give the explicit
functional form of the function $n=n_0\phi$, which may to determined from the solution (\ref{NLsolution}) of the
stationarity equation of the variational problem for the effective potential.
The free parameters left in the variables of the scaling function $f(x,y)$ correspond to the choice of units of these variables, whereas the
objective information is contained in the form of the scaling function~\cite{Vasiliev04,Adzhemyan99}.

From the explicit solution (\ref{NLsolution}) we obtain the generic expression
\begin{multline}
\label{eq:ExpansionOfScaling}
h(x,y)={1\over f(x,y)}\\
=1+2xy\left\{1+{\displaystyle{{\lambda}^* }\over \displaystyle 4\pi}\left[1-\gamma
-\ln\left(2u^*x\right)\right]\right\}\,,
\end{multline}
the substitution in which of the various fixed-point values $\lambda^* $ (at the leading order $\lambda^*\approx 2\pi\overline{\lambda}^*$)
and $u^*$ in the leading approximation
together with the substitution of $x=\nu\mu^2\tau$ and $y=\tilde{C}\overline{y}$ from
(\ref{eq:initialdensityasy}) and (\ref{eq:defalpha})
yields the corresponding $\varepsilon$, $\Delta$ expansions
of the asymptotic expression of the normalized number density.

Below, we list the scaling functions  $h(x,y)$ and the dynamic exponents $\alpha$ at the stable fixed points
in the next-to-leading-order approximation.\\
({\it i}$\,$) At the trivial (Gaussian) fixed point (\ref{Gaussian})
the mean-field behavior takes place with
\begin{align}
\label{eq:hGaussian}
h(x,y)&=1+2xy\,,\nonumber\\*
\alpha&=1\,.
\end{align}
({\it ii}$\,$)
The thermal (short-range) fixed point
(\ref{SR}) leads to scaling function and decay exponent
\begin{align}
\label{eq:hSR}
h(x,y)&=1+2xy\left\{1-{\displaystyle{\Delta }\over \displaystyle 2}\left[1-\gamma
-\ln \left(\sqrt{17}-1\right)\,x\right]\right\}
\,,\nonumber\\*
\alpha&=1+{\Delta\over 2}+{\Delta^2\over 2}\,.
\end{align}
Here, the last coefficient is actually a result of numerical calculation, which in the standard accuracy of Mathematica
is equal to 0.5. We have not been able to sort out this result analytically, but think that most probably the coefficient of
the $\Delta^2$ term in the decay exponent $\alpha$ in (\ref{eq:hSR}) really is ${1\over 2}$.\\
({\it iii}$\,$)
The kinetic fixed point with an anomalous reaction rate
(\ref{K1}) corresponds to
\begin{eqnarray}
\label{eq:hK1}
h(x,y)&=&1+2xy \biggl\{ 1-{\displaystyle{\epsilon+3\Delta }\over \displaystyle 3}\times \nonumber\\
& & \left[1-\gamma-\ln \left(\sqrt{17}-1\right)\,x\right]\biggl\}
\,,\nonumber\\*
\alpha&=&1+{3\Delta+\epsilon\over 3-\epsilon}\,,
\end{eqnarray}
with an exact value of the decay exponent. \\
({\it iv}$\,$)
At the kinetic fixed point with mean-field-like reaction rate
(\ref{K2}) we obtain
\begin{align}
\label{eq:hK2}
h(x,y)&=1+2xy
\,,\nonumber\\*
\alpha&=1\,.
\end{align}
In the actual asymptotic expression corresponding to (\ref{eq:scalingf}) the argument $y\to \widetilde{C}_n\overline{y}$
is different from that of the Gaussian fixed point.

\subsection{Asymptotic behavior in two dimensions}
To complete the picture, we recapitulate -- with a little bit more detail --
the asymptotic behavior of the number density in the physical space dimension $d=2$ predicted
within the present approach \cite{Hnatic} (it turns out that for these conclusions the one-loop
calculation is sufficient).
On the ray $\varepsilon\le 0$, $\Delta=0$ logarithmic corrections to the mean-field decay take place. The integral
determining the asymptotic behavior of the variable (\ref{eq:initialdensity}) yields in this case
\begin{equation}
\label{eq:asyamplitudelog}
{ \int_\tau^t{\displaystyle \gamma_4 ds\over\displaystyle  (2-\gamma_1)s}}
\operatornamewithlimits{\sim}_{t\to\infty}-{1\over 2}\,\ln\ln\left({t\over \tau}\right)+\tilde{c}_4(\tau)\,,
\end{equation}
with corrections vanishing in the limit $t\to\infty$. Therefore, in the vicinity of the fixed point
\begin{equation}
\label{eq:initialdensityasylog}
\overline{\lambda}\overline{u}\,{\overline{n}_0\over \mu^d}
\sim \lambda u\,{n_0\over \mu^d} \left({t\over \tau}\right)\ln^{-1/2}\left({t\over \tau}\right)\widetilde{C}_n
\equiv \overline{y}\,\widetilde{C}_n\,.
\end{equation}
The scaling function $h$ is of the simple form
\begin{equation}
\label{eq:hlog}
h(x,y)=1+2xy\nonumber
\end{equation}
and gives rise to asymptotic decay slower than in the mean-field case by a logarithmic factor:
\[
{n}\sim {\ln^{1/2}\left(t/ \tau\right)\over 2\nu\lambda u \widetilde{C}_n t}\,.
\]
It is worth noting that this logarithmic slowing down is weaker than that brought about the density fluctuations only \cite{Peliti}
and this change is produced even by the ubiquitous thermal fluctuations of the fluid, when the reaction is taking
place in gaseous or liquid media.

On the open ray $\varepsilon> 0$, $\Delta=0$ the kinetic fixed point with mean-field-like reaction rate (\ref{K2}) is
stable and the asymptotic behavior is given by (\ref{eq:hK2}) regardless of the value of the falloff exponent of the
random forcing in the Navier-Stokes equation. In particular, only the amplitude factor in
the asymptotic decay rate in two dimensions is affected by the developed turbulent flow with Kolmogorov scaling, which
corresponds to the value $\varepsilon=2$.
This is in accord with the results obtained in the case of quenched solenoidal flow with long-range correlations
\cite{Deem2,Tran} as
well as with the usual picture of having the maximal reaction rate in a well-mixed system.
\section{Conclusion}

In conclusion, we have analyzed the effect of density and velocity
fluctuations on the reaction kinetics of
the single-species decay $A+A\to \emptyset$ universality class in the
framework of field-theoretic renormalization group and calculated the
scaling function and the decay exponent of the mean number density for the four asymptotic patterns
predicted by the RG.

We have calculated the relevant renormalization constants at two-loop level
and found the decay exponent of the mean number density at this order
of the $\epsilon$, $\Delta$ expansion for four IR stable fixed points of the
RG, whose regions of stability cover the whole parametric space in the vicinity
of the origin in the $\epsilon$, $\Delta$ plane. The decay exponent assumes
the mean-field value in the basins of attraction of the trivial fixed point (\ref{Gaussian}) and
of the kinetic fixed point (\ref{K2}) with dominant fluctuations of the random force
of the Navier-Stokes equation. At the kinetic fixed point with finite rate coefficient (\ref{K1}) the
value of the decay exponent is determined exactly by the fixed-point equations. At the thermal (short-range)
fixed point (\ref{SR}) the decay exponent possesses a non-trivial $\epsilon$, $\Delta$ expansion.
We have calculated three first terms of this expansion
and thus pushed its accuracy by one order further than in
in the one-loop calculation as well as in the case of kinetic fixed point with
anomalous reaction rate. We have also found an overlap between the regions
of stability of these two asymptotic patterns, which means that the asymptotic
behavior of system depends on the details of the approach to the fixed point.

Using a variational approach, we have inferred a renormalized
fluctuation-amended rate equation with the account of one-loop
corrections,
which again is a step forward in the accuracy of description
of the system and opens the possibility to analyze direct effects of
on the number density
of the velocity fluctuations, which do not enter explicitly in the
leading-order mean-field rate equation.
This non-linear integro - differential equation has been
solved iteratively in the framework of the $\epsilon$, $\Delta$
expansion and the scaling function for the mean number density has
been calculated for the four IR stable regimes. The scaling function
assumes the mean-field form (exactly) in the basins of attraction of
the trivial fixed point and the kinetic fixed point with dominant
fluctuations of the random force. At the kinetic fixed point with
finite rate coefficient  and at the thermal fixed point the scaling
function possesses a non-trivial $\epsilon$, $\Delta$ expansion,
which we have calculated at the linear order.

Fluctuations of the
random advection field affect heavily the long-time asymptotic
behavior of the system: the kinetic fixed points are brought about
by the velocity fluctuations as well as the non-trivial series
expansion of the decay exponent at the thermal fixed point (without
velocity fluctuations, the decay exponent is fixed to the one-loop
value, because there are no high-order corrections to the rate
constant in this case). Predictions of the renormalization-group
analysis for the reaction $A+A\to \emptyset$ in quenched random
fields have been corroborated by numerical simulations
\cite{Chung,Tran}. In the case of dynamically generated random drift
this is seems to be a much more demanding task, but would surely be
highly desirable, since the experimental data for reaction processes
is quite scarce \cite{Tauber05}.

The integro-differential equation for the calculation of the concentration
proposed here and solved analytically in a suitable approximation allows for
heterogeneous solutions as well. Since the heterogeneous concentration most
probably has to be found numerically, analysis of this issue is beyond the scope
of the present paper. We intend to carry out this investigation in the near future,
all the more so because there is significant recent interest in the effect of
random drift on nonlinear diffusion equations \cite{Tel05,Benzi12}, to which the stationarity
equations of the proposed variational functional belong. However, most of these
problems include competition between growth and decline of population and give
rise to more complicated models than that at hand. The potential of the
field-theoretic approach in analytic investigation of stochastic problems is, however, far from
being exhausted, therefore in the future we hope to deal with the diffusion-limited birth-death
processes in random flows within this framework as well.

\begin{acknowledgement}
The work was supported by VEGA grant 0173 of Slovak A\-ca\-de\-my of Sciences, and by Centre of Excellency for
Nanofluid of IEP SAS. This article was also created by implementation
of the Cooperative phenomena and phase transitions
in nanosystems with perspective utilization in nano- and
biotechnology projects No 26220120021 and No 26220120033. Funding for the operational
research and development program was provided by
the European Regional Development Fund. T.L. was sponsored by a scholarship
grant by the Aktion \"Osterreich-Slowakei.
\end{acknowledgement}


\appendix

\section{Explicit form of the renormalization constants}
\label{app:ABcoeff}

The coefficient functions $A_{ij}$ and $B_{ij}$ of the renormalization constant (\ref{eq:Z2}) are
\begin{align*}
 A_{11} &= -\frac{(1+\xi)u^2+(3\xi+2)u+6\xi+1}{512u(1+u)^3\xi},\\
A_{12}&= -\frac{(1+\xi)u^2+(1+3\xi)u+6\xi-4}{256u(1+u)^3(1-\xi)},\\
A_{22}&=-\frac{u+5}{512u(1+u)^3},\\
B_{11} &=B_1(u)+ B_2(u,\xi),\\
B_{12} &= 4[B_1(u)+ B_3(u,\xi)],\\
B_{22}& = -B_1(u)-B_2(u,-1),
 \end{align*}
where the functions $B_1,B_2$ and $B_3$ are given as

 \begin{align*}
 \begin{split}
 &B_1(u) = \frac{1}{1024u^4(1+u)^3(u-1)}\biggl[\\
 &-12u^3(1+u)\ln\frac{2}{1+u}
 +32u^4(1+u)^2\ln 2\\
 &+
2u^3(1+u)^2(u+10)\ln\frac{4}{3}
-32u^3(1+u)^3{\rm artanh}\frac{1}{2}\\
  &+ 4(2u^6+6u^5+7u^4+10u^3-4u-1)\ln\frac{1+2u}{(1+u)^2}\\
&-4u^3(1+u)
[4u^2+20u+9]\ln\frac{1+2u}{2+2u}\\
&+16u^3(1+u)^3{\rm artanh}\frac{u}{u+1}\\
&+
8u^3(1+u)^2(u-1)(\gamma+\psi(3/2))\\
&-u^2(23u^4+38u^3+17u^2+22u+4)\biggl]\\
&-
 \frac{1}{128\pi u(1+u)^2(u-1)}\int_1^\infty dq\int_{-1}^1 dz F(q,z,u)\,,
 \end{split}\\
 &B_2(u,\xi)  =\xi\frac{(2+4\xi)u^2+(10\xi+8)u+38+22\xi}{1024u(1+u)^3(2+\xi)},\\
 &B_3(u,\xi)  = -\frac{(8\xi+4)u^2+(14\xi+10)u+22-10\xi}{1024u(1+u)^3}\\
\end{align*}
with the function $F$ given by the expression
\begin{align*}
&F(x,z,u)  =  (1-z^2)^{1/2}\frac{M(x,z,u)}{N(x,z,u)} \\
\begin{split}
&M(x,z,u)  =  (x^6+1)[z^3(24u^3+24u^2+72u+72)\\
&-z(8u^3+12u^2+8u+60)]+
(x^5+x)[-z^4(40u^3\\
&+88u^2+120u+
264)+ z^2(-4u^3+16u^2+108u+168)\\
&+4u^3+14u^2+28u+18]+
(x^4+x^2)[z^5(16u^3+96u^2\\
&+48u+
288)+  z^3(12u^4+64u^2+ 128u^2+96u+180)\\
&-z(4u^4+26u^3+92u^2+174u+312)] + x^3[-z^6(32u^2\\ &
+96)- z^4(8u^4+64u^3+240u^2+144u+600)\\
&+ z^2(-8u^4+4u^3+84u^2+108u+452)+2u^4+ 6u^3\\&
+26u^2+58u+36],
\end{split}
\\
&N(x,z,u) = (1+x^2-2xz)(1+x^2-xz)\\
&\times((1+u)x^2+2-2xz)(1+u+2x^2-2xz)\,.
\end{align*}
The coefficient function $C$ of the renormalization constant (\ref{eq:Z4}) is
\begin{equation}
\begin{split}
 C(u,\xi)&=
  - \frac{1}{8u\pi } \int_{-1}^1 dz (1-z^2)^{\frac{1}{2}} G(z,u)\\
&+  \frac{1}{8 u(1+u)}\biggl(\ln\frac{1+u}{2u}+1+\frac{2+u}{u}\ln\frac{u+2}{2u+2}+\xi\biggl),
  \end{split}
\nonumber
\end{equation}
\text{where}
\begin{equation}
\begin{split}
&G(z,u) =  \frac{4}{(1-u)^2+4uz^2} \biggl\{\frac{u-1}{2}\ln\frac{2u}{1+u}\\
&-
\frac{2(1+u)z}{\sqrt{1-z^2}}\biggl[\frac{\pi}{2}-\arctan\sqrt{\frac{1+z}{1-z}} \biggl]\nonumber\\
& +\frac{u(u+3)z}{\sqrt{2u(1+u)-u^2z^2}}\biggl[\pi-\arctan\frac{zu+u+1}{\sqrt{2u(1+u)-u^2z^2}}\\
&-\arctan\frac{(2+z)u}{\sqrt{2u(1+u)-u^2z^2}} \biggl]\biggl\}\,.
  \end{split}
\end{equation}

\section{Fixed points}
\label{app:fixedpoints}
The coefficient functions of the fixed points (\ref{K1}) and (\ref{K2}) are
\begin{align*}
\begin{split}
 &u_1^*(\xi)=\frac{8R}{3\sqrt{17}}
 -\frac{8192}{3\sqrt{17}}B_1(u^*_0)-
  \frac{1}{432\sqrt{17}(1+\xi)^2(2+\xi)}\\
  &\times\biggl[
  (21384-648\sqrt{17})\xi^4 \\
 & +(52512-2592\sqrt{17}) \xi^3+(22192-2736\sqrt{17})\xi^2\\
  &+(72\sqrt{17}-29064)\xi+720\sqrt{17}-18768
  \biggl]\,,
\end{split}
  \\
\begin{split}
 & g_{12}^*(\xi)=64\frac{R(2+3\xi)-1}{27(1+\xi)}\\
 &-16\frac{2+3\xi}{243(1+\xi)^4(2+\xi)}\biggl[
   45\xi^4+213\xi^3+349\xi^2+231\xi+50
   \biggl]\,,
\end{split}
   \\
 \begin{split}
 &  g_{22}^*(\xi)=64\frac{1+R}{27(1+\xi)}\\
 &-16\frac{2+3\xi}{243(1+\xi)^4(2+\xi)}\biggl[
  57\xi^4+171\xi^3+185\xi^2+93\xi+22
   \biggl]\,.
 \end{split}
\end{align*}


\end{document}